\documentclass[journal]{IEEEtran}
\usepackage{stfloats}
\usepackage{amsmath}
\usepackage{epsfig}
\usepackage{amssymb}
\usepackage{verbatim}
\usepackage{delarray}
\usepackage{graphics}
\usepackage{epic}

\newcommand{\lra}{\leftrightarrow}
\newcommand{\inner}[2]{\langle{#1},{#2}\rangle}

\newcommand{\A}{{\mathcal{A}}}

\newcommand{\CC}{{\mathcal{C}}}
\newcommand{\D}{{\mathcal{D}}}

\newcommand{\FF}{{\mathcal{F}}}
\newcommand{\G}{{\mathcal{G}}}
\newcommand{\HH}{{\mathcal{H}}}
\newcommand{\I}{{\mathcal{I}}}
\newcommand{\J}{{\mathcal{J}}}
\newcommand{\K}{{\mathcal{K}}}

\newcommand{\PP}{{\mathcal{P}}}
\newcommand{\SSS}{{\mathcal{S}}}
\newcommand{\T}{{\mathcal{T}}}

\newcommand{\F}{{\mathbb{F}}}
\newcommand{\R}{{\mathbb{R}}}
\newcommand{\Z}{{\mathbb{Z}}}
\newcommand{\zerob}{{\mathbf 0}}
\newcommand{\oneb}{{\mathbf 1}}
\newcommand{\ab}{{\mathbf a}}

\newcommand{\gb}{{\mathbf g}}
\newcommand{\hb}{{\mathbf h}}

\newcommand{\rb}{{\mathbf r}}
\renewcommand{\sb}{{\mathbf s}}

\newcommand{\alphab}{\mbox{\boldmath$\alpha$}}

\newcommand{\Bf}{{\mathfrak{B}}}

\newcommand{\ie}{{\em i.e., }}
\newcommand{\eg}{{\em e.g., }}
\newcommand{\cf}{\emph{cf.\ }}

\newcommand{\openbox}{\leavevmode
     \hbox to.77778em{%
     \hfil\vrule
     \vbox to.675em{\hrule width.6em\vfil\hrule}%
     \vrule\hfil}}
\newcommand{\qed}{\hspace*{1cm}\hspace*{\fill}\openbox}

\begin{document}

\title{Codes on Graphs: \\  Observability, Controllability and Local Reducibility}

\author{G. David Forney, Jr.\ and Heide Gluesing-Luerssen

\thanks{
The authors are with the
Laboratory for Information and Decision Systems,
Massachusetts Institute of Technology,
Cambridge, MA 02139 (email: forneyd@comcast.net) and Department of Mathematics, University of Kentucky, Lexington KY 40506-0027 (email: heide.gl@uky.edu), respectively.
H. Gluesing-Luerssen was partially supported by the National Science Foundation Grant DMS-0908379. 
This paper was presented in part at the 2012 Information Theory and Applications Workshop, San Diego, CA, and at the 2012 IEEE International Symposium on Information Theory, Cambridge, MA.}}
\date{}

\maketitle

\begin{abstract}
This paper investigates properties of realizations of linear or group codes on general graphs that lead to local reducibility.

Trimness and properness are dual properties of constraint codes.  A linear or group realization with a constraint code that is not both trim and proper is locally reducible.  A linear or group realization on a finite cycle-free graph is minimal if and only if every local constraint code is trim and proper.    

A realization is called observable if there is a one-to-one correspondence between codewords and configurations, and controllable if it has independent constraints.  A linear or group realization is observable if and only if its dual is controllable.  A simple counting test for controllability is given.   An unobservable or uncontrollable realization is locally reducible.  Parity-check realizations are controllable if and only if they have independent parity checks.  In an uncontrollable tail-biting trellis realization, the behavior partitions into disconnected subbehaviors, but this property does not hold for non-trellis realizations.  On a general graph, the support of an unobservable configuration is a generalized cycle.  \end{abstract}

\vspace{-2ex}
\section{Introduction}

The theory of minimal realizations of linear codes and systems on a conventional discrete time axis has long been well understood.  In classical minimal realization theory, the concepts of observability and controllability are central;  indeed, the classical mantra is  ``minimal = controllable + observable." 

Moreover, as is well known, classical minimal realization theory extends straightforwardly to realizations of linear (or abelian group) codes or systems on cycle-free graphs  \cite{F01}.  Given such a graph, the State Space Theorem defines a unique minimal ``state space" for every internal (``state") variable.  

On the other hand, powerful codes that can approach capacity with iterative decoding (\eg low-density parity-check (LDPC) or turbo codes) are necessarily defined on graphs that have cycles.   Progress in extending minimal realization theory to general graphs with cycles has been slow. (Also, it is not clear that minimal realizations are necessarily the best ones for iterative decoding;  see Section \ref{Section 4.8}.)

A fundamental issue is that in general there is no unique minimal realization of a linear code or system on a graph that contains cycles.  As was observed by Koetter and Vardy \cite{KV03}, even for the simplest such graph comprising a single cycle (called a ``tail-biting trellis realization"), there are various notions of minimality, and for any such notion there is only a partial ordering of tail-biting trellis realizations.  However, Koetter and Vardy were able to specify a small, tractable set of possibly minimal tail-biting trellises, by showing that every linear tail-biting trellis realization factors into a product of one-dimensional ``atomic" realizations \cite{KV02} (as do conventional linear trellis realizations \cite{KS95}).

On more general graphs, there is no unique minimal realization, and also no such product factorization (see Appendix A).  How then can we pursue something like minimality?

The main idea of this paper is to look for possibilities for ``locally" reducing the complexity of a realization--- \eg by reducing the size of one internal variable alphabet (``state space")--- without changing the code that is realized or the complexity of the rest of the realization.  Such an approach follows a long tradition in the conventional trellis minimization literature (\eg \cite{LFMT, V98}), and was previously explored by Koetter \cite{K02}.  In Appendix B, we give a brief summary of Koetter's pioneering work in the language of this paper.

For instance, a realization in which some state variable value never occurs  is obviously locally reducible, as the state alphabet may be trimmed to the set of values that actually occur.
But as Koetter \cite{K02} observed, duality considerations using the ``dual realizations" of \cite{F01} can reveal less obvious dual local reducibility criteria.  For example, we give a simple proof that a constraint code is not (state-)trim if and only if the dual constraint code is not  ``proper,"  in a sense that extends the traditional definition of ``properness" for trellis realizations \cite{V98}.  By duality, any realization that involves an ``improper" constraint code must therefore be locally reducible.  

We next show that a finite linear or group normal realization on a cycle-free graph is minimal if and only if every constraint code is both trim and proper.  Although this simple result has long been known for conventional trellis realizations \cite{V98}, it seems to be new for realizations on more general cycle-free graphs.  This development yields a constructive proof of the fundamental State Space Theorem, and leads immediately to straightforward iterative minimization algorithms.

Next, we consider linear or group realizations on general graphs that are not one-to-one--- \ie every codeword in $\CC$ is realized by multiple configurations. Such realizations are called \emph{unobservable}, since the values of the internal variables are not determined by those of the external variables.  Such realizations would seem to be obviously undesirable for iterative decoding;  see Section \ref{Section 4.8}.  

We show that any unobservable realization is locally reducible.  Therefore we may always assume that any linear or group realization on a general graph is observable.

Following the principle that the dual of a locally reducible normal realization of $\CC$ is  locally reducible, we show that the dual of an unobservable linear realization is a realization with dependent constraints.  For consistency with classical terminology, we call such a realization \emph{uncontrollable}.  By duality, an uncontrollable realization is locally reducible.  

However, it is not so clear that an uncontrollable realization is unsuitable for iterative decoding;  see Section \ref{Section 4.8}.  For example, we show that any parity-check realization with redundant parity checks is uncontrollable;  nevertheless, LDPC codes with redundant checks have sometimes been preferred in practice.

We show that an uncontrollable  tail-biting trellis realization consists of disjoint subrealizations, as with classical uncontrollable conventional trellis realizations.  However, realizations on general graphs do not in general have this property.

Finally, given an unobservable realization on a general graph, the support of the unobservable configurations must be a cycle or generalized cycle (a subgraph whose vertices all have degree 2 or more).  In the dual uncontrollable realization, there exist nonlocal constraints that are defined on the same support.  In the binary case, this support must actually be an Eulerian cycle (a generalized cycle whose vertices all have even degree).  

We may summarize our progress on establishing conditions for local reducibility as follows.  We partition the universe of linear or group codes on finite graphs into four quadrants, as shown in the $2 \times 2$ matrix below, where one axis distinguishes trellis from non-trellis graphs, and one cycle-free from cyclic. 

{
\setlength{\unitlength}{4pt}
\centering
\begin{picture}(30,17)(-15, -1)
\put(-10,2){\small cyclic}
\put(-10,7){\small cycle-free}
\put(5,12){\small trellis}
\put(25,12){\small non-trellis}
\put(0,0){\line(1,0){40}}
\put(0,5){\line(1,0){40}}
\put(0,10){\line(1,0){40}}
\put(0,0){\line(0,1){10}}
\put(20,0){\line(0,1){10}}
\put(40,0){\line(0,1){10}}
\put(0,0){\framebox(20,5){\small tail-biting  trellises}}
\put(0,5){\framebox(20,5){\small conventional trellises}}
\put(20,0){\framebox(20,5){\small general graphs}}
\put(20,5){\framebox(20,5){\small cycle-free graphs}}
\end{picture}
}

For conventional trellis realizations, simple conditions for minimality have long been known.  In this paper, we find a simple, conclusive, and locally testable condition for minimality for a general finite linear or group cycle-free realization: 
\begin{center}
minimal $\Leftrightarrow$ every constraint code is trim and proper.  
\end{center}

For linear tail-biting trellis realizations, which in general have no unique minimal realization, we make some progress in this paper toward finding conditions for local irreducibility.
In a subsequent paper \cite{GLF}, we will give necessary and sufficient conditions for a linear tail-biting trellis realization to be irreducible, under a more refined definition of local reducibility.  These results largely settle the tail-biting trellis case, although we consider that there are still some open issues in this case.

Finally, for the general case, beyond trimness and properness, the dual concepts of observability and controllability yield further local reducibility criteria;  however, this case remains largely open.

\pagebreak
\section{Codes, realizations, and duality}\label{Section 2}

In this paper we will focus on linear codes over a finite field $\F$.  Everything generalizes to group codes over finite abelian groups, and to linear systems over the real or complex field.  We will state some of our most basic results for both the linear and group cases, since the linear case is often more familiar, whereas the group-theoretic statements and proofs  are often more transparent.

\subsection{Codes and realizations}

A linear code $\CC$ over a finite field $\F$ is a linear subspace of a \emph{symbol configuration space} $\A = \Pi_{k \in \I_\A} \A_k$, where each \emph{symbol alphabet} $\A_k$ is a finite-dimensional vector space over $\F$, and $\I_\A$ is a discrete index set.  We will usually assume that $\I_\A$ is finite (\ie $\CC$ is a block code);  then the symbol configuration space $\A$ is finite.  We may alternatively refer to a symbol variable as an \emph{external variable} $A_k$, which takes values $a_k$ in an external variable alphabet $\A_k$.

Similarly, a group code $\CC$ is a subgroup of a symbol configuration space $\A = \Pi_{k \in \I_\A} \A_k$, where each symbol alphabet $\A_k$ is a finite abelian group.  If $\I_\A$ is finite, then $|\A| = \Pi_{k \in \I_\A} |\A_k|$ is finite.

For a \emph{state realization} of $\CC$, we define also a set $\{\SSS_j: j \in \I_\SSS\}$ of \emph{state spaces} $\SSS_j$ indexed by a state index set $\I_\SSS$, and a set $\{\CC_i: i \in \I_\CC\}$  of local \emph{constraint codes} $\CC_i$ indexed by a constraint index set $\I_\CC$, where each constraint code $\CC_i$ involves subsets  of the symbol and state variables indexed by  $\I_{\A^{(i)}} \subseteq \I_\A$ and $ \I_{\SSS^{(i)}} \subseteq \I_\SSS$, so $\CC_i \subseteq \A^{(i)} \times \SSS^{(i)}$, where $\A^{(i)} = \prod_{k \in \I_{\A^{(i)}}} \A_k$ and $\SSS^{(i)} = \prod_{j \in \I_{\SSS^{(i)}}} \SSS_j$.  

 In a linear realization, each state space $\SSS_j$ and each constraint code $\CC_i$ is a vector space over $\F$;  in a group realization, each state space and constraint code is a finite abelian group.  
 
 The \emph{state configuration space} is $\SSS = \Pi_{j \in \I_\SSS} \SSS_j$.     We may alternatively refer to a state variable as an \emph{internal variable} $S_j$, which takes values $s_j$ in an internal variable alphabet $\SSS_j$.

The \emph{degree} of a variable is the number of constraint codes in which it is involved; the \emph{degree} of a constraint code is the number of variables that it involves.

A \emph{configuration} (or  \emph{trajectory}) of the realization is a pair $(\ab, \sb) \in \A \times \SSS$ for which all constraints are satisfied;  \ie $(\ab^{(i)}, \sb^{(i)}) \in \CC_i, \forall i \in \I_\CC$.  Its \emph{full behavior}  is the set $\Bf$ of all such valid configurations. The \emph{code} $\CC$ realized by the realization is then the set of all $\ab \in \A$ that appear in some $(\ab, \sb) \in \Bf$;  \ie $\CC$ is the projection of $\Bf$ onto $\A$.  A linear realization realizes a linear code;  a group realization realizes a group code.

\subsection{Observability}

A state realization is called \emph{one-to-one}, or \emph{observable}, if there is precisely one pair $(\ab, \sb) \in \Bf$ corresponding to each $\ab \in \CC$. 

For a linear or group realization, we define the \emph{unobservable state configuration space} $\SSS^u$  as the subspace of \emph{unobservable state configurations} $\sb \in \SSS$ such that $(\zerob, \sb) \in \Bf$;  thus such a realization is observable if and only if $\SSS^u$ is trivial. Indeed, $\SSS^u$ is evidently isomorphic to the kernel $\K^u = \{(\zerob, \sb) \in \Bf\}$ of the projection of $\Bf$ onto $\A$, whose image is $\CC$, so by the fundamental theorem of homomorphisms $\CC \cong \Bf/\K^u$.  Therefore $|\CC| = |\Bf|/|\SSS^u|$, or, in the linear case, $\dim \CC = \dim \Bf - \dim \SSS^u$.

\subsection{Minimality and local reducibility}

As in~\cite{KV03}, a realization with state spaces~$\SSS_j$ will be called \emph{minimal} if there exists no realization of the same code with state spaces~$\tilde{\SSS}_j$ that has the same graph topology (\ie the same index sets~$\I_\A, \I_\SSS,\,\I_\CC, \I_{\A^{(i)}},\,\I_{\SSS^{(i)}}$) such that
$|\tilde{\SSS}_j| \leq |\SSS_j|$ for all~$j$, with at least one strict inequality.

A realization of a code~$\CC$ with state spaces~$\SSS_j$ and constraint codes~$\CC_i$ will be called \emph{locally reducible}
if it is possible to replace a single state space~$\SSS_j$ by a strictly smaller space, and the constraint codes involving~$\SSS_j$ by constraint codes that are no bigger, without changing the rest of the realization or the code that it realizes.  Evidently a locally reducible realization is nonminimal.

Our basic tools for local reduction will be trimming (restricting) and merging (taking quotients) of state spaces, which we will see are dual operations.

\subsection{State variables of degree one}

As a first exercise in local reducibility, we observe that any realization that includes a nontrivial state variable of degree 1 is locally reducible, since we may always reduce the alphabet of that variable  to the trivial alphabet (\ie merge all states with the zero state) without changing the code realized by the realization, there being no other constraints on that state variable.  Thus we will assume henceforth that state variables of degree 1 have a trivial state space.  

We note further that, in any state realization, a state variable whose alphabet size is one may optionally be deleted from the realization without affecting the code that is realized.    However, this is not the usual convention for end states in trellis realizations;  see below.

\subsection{Normal realizations}

A state realization is called \emph{normal} if the degree of every symbol variable is 1, and the degree of every state variable is 2. (More generally, a state variable could have degree 1, provided that its alphabet is trivial;  see the previous subsection.)

As shown in \cite{F01}, any realization may be straightforwardly ``normalized" by introducing replica internal variables and equality constraints, as follows.  For every constraint code $\CC_i$ and every variable $A_k$ or $S_j$ involved in $\CC_i$, create a replica internal variable $A_{ki}$ or $S_{ji}$;  then, for every $S_j$, introduce an equality constraint on all the replicas $S_{ji}$ of $S_j$ and delete the original variable $S_j$, and for every $A_k$, introduce an equality constraint on all the replicas $A_{ki}$ of $A_k$, plus the original external variable $A_k$.  The resulting realization is normal, realizes the same code $\CC$, and has essentially the same graph topology.  Thus we will assume henceforth without essential loss of generality that all realizations are normal.

A normal realization has a natural graphical representation, called a \emph{normal graph}, in which constraints are represented by vertices, internal variables by ordinary edges (edges of degree 2), and external variables by half-edges (edges of degree 1).  An edge or half-edge is incident on a vertex whenever the corresponding variable is involved in the corresponding constraint code.

If the graph of a realization is disconnected, then the behavior and the code that it realizes are the Cartesian products of the respective behaviors and codes realized by the components.  We will therefore assume that the graphs of all realizations are connected.

\subsection{Trellis realizations}

A \emph{trellis realization} is a state realization in which every constraint code involves precisely two state variables, every state variable is involved in one or two constraint codes and every symbol variable is involved in one constraint code.  Thus the graph of a trellis realization must be one of the following (if connected):
\begin{itemize}
\item a finite, semi-infinite or bi-infinite chain graph (called a ``path'' in graph theory), which we call a \emph{conventional trellis realization};
\item a single-cycle graph, which we call a \emph{tail-biting trellis realization}.
\end{itemize}
Thus bi-infinite and tail-biting trellis realizations are inherently normal.  Finite or semi-infinite conventional trellis realizations are not strictly normal, because they have ``end" state variables of degree 1, but they may be taken to be normal if the associated degree-1 state alphabets are  trivial.

We note that a finite conventional trellis realization may be considered to be a special case of a tail-biting trellis realization in which one state variable has a trivial alphabet, and therefore could be deleted.

We will sometimes depict a trellis realization by a traditional \emph{trellis diagram}, in which all branches, states and symbols are depicted explicitly.

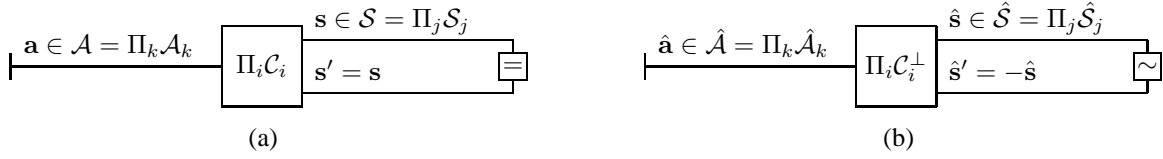
\begin{figure*}[!t]
\setlength{\unitlength}{5pt}
\centering
\begin{picture}(70,8)(0, -2)
\put(-8,2){\line(0,1){2}}
\put(-8,3){\line(1,0){16}}
\put(-7,4){$\ab \in \A = \Pi_k \A_k$}
\put(8,0){\framebox(6,6){$\Pi_i \CC_i$}}
\put(14,5){\line(1,0){16}}
\put(14,1){\line(1,0){16}}
\put(15,6){$\sb \in \SSS = \Pi_j \SSS_j$}
\put(15,2){$\sb' = \sb$}
\put(30,5){\line(0,-1){1}}
\put(30,1){\line(0,1){1}}
\put(29,2){\framebox(2,2){$=$}}
\put(10,-3){(a)}
\put(40,2){\line(0,1){2}}
\put(40,3){\line(1,0){16}}
\put(41,4){$\hat{\ab} \in \hat{\A} = \Pi_k \hat{\A}_k$}
\put(56,0){\framebox(6,6){$\Pi_i \CC_i^\perp$}}
\put(62,5){\line(1,0){16}}
\put(62,1){\line(1,0){16}}
\put(63,6){$\hat{\sb} \in \hat{\SSS} = \Pi_j \hat{\SSS}_j$}
\put(63,2){$\hat{\sb}' = -\hat{\sb}$}
\put(78,5){\line(0,-1){1}}
\put(78,1){\line(0,1){1}}
\put(77,2){\framebox(2,2){$\sim$}}
\put(58,-3){(b)}
\end{picture}
\caption{Normal realization duality:  Dual realizations realize orthogonal codes.}
\label{NGDT}
\end{figure*}

\subsection{Duality}

We now briefly recapitulate some of the basic duality principles for both the group and linear cases that will be used heavily in this paper, following \cite{FT04}.

If $G$ is a finite abelian group, then its \emph{dual group} $\hat{G}$ is the character group of $G$, namely the set of all homomorphisms from $G$ to the circle group $\R/\Z$.  The dual group  of $\hat{G}$ may be taken as $G$.  In the finite abelian case, $G$ and $\hat{G}$ are isomorphic.

  For $g \in G$ and $\hat{g} \in \hat{G}$, there is a well-defined \emph{pairing} $\inner{\hat{g}}{g} = \hat{g}(g) \in \R/\Z$ that has the usual bihomomorphic properties:  \eg $\inner{0}{g} = \inner{\hat{g}}{0} = 0$,  $\inner{\hat{g}_1 + \hat{g}_2}{g} = \inner{\hat{g}_1}{g} + \inner{\hat{g}_2}{g}$, and so forth.
  
  Elements $g \in G$ and $\hat{g} \in \hat{G}$ are said to be \emph{orthogonal} if $\inner{\hat{g}}{g} = 0$.  If $H$ is a subgroup of $G$, then the orthogonal subgroup $H^\perp$ is the set of all elements of $\hat{G}$ that are orthogonal to all elements of $H$.  $H^\perp$ is indeed a subgroup of $\hat{G}$, and the orthogonal subgroup to $H^\perp$ is $H$.
  
  The dual group $\hat{H}$ to $H \subseteq G$ is isomorphic to the quotient group $\hat{G}/H^\perp$.  In the finite abelian case, this implies $|H| |H^\perp| = |G| = |\hat{G}|$.
  
  Similarly, if $G$ is a finite-dimensional vector space over a finite field $\F$, then it has a \emph{dual space}  $\hat{G}$, a vector space over $\F$ of the same dimension, which is furnished with an \emph{inner product} $\inner{\hat{g}}{g}   \in \F$ that is a bilinear form:  \ie $\inner{0}{g} = \inner{\hat{g}}{0} = 0$,  $\inner{\hat{g}_1 + \hat{g}_2}{g} = \inner{\hat{g}_1}{g} + \inner{\hat{g}_2}{g}$, and so forth.  It is well known that all dual spaces of~$G$ are isomorphic, where the isomorphism preserves the inner product.
If~$G=\F^\ell$ for some~$\ell$, then we may and will choose $\hat{G}=\F^\ell$, along with the standard inner (dot) product on~$\F^\ell$.
  
  Moreover, if $\{b_1, \ldots, b_n\}$ is a basis for an $n$-dimensional vector space $G$, then there exists a dual basis $\{\hat{b}_1, \ldots, \hat{b}_n\}$ for $\hat{G}$ such that $\inner{\hat{b}_i}{b_j} = \delta_{ij}$, the Kronecker delta function.  Given any $g \in G$, there exists a unique \emph{coordinate vector} $\alphab \in \F^n$ such that $g = \sum_j \alpha_j b_j$; this establishes an isomorphism between $G$ and $\F^n$.  Similarly, each $\hat{g} \in \hat{G}$ has a unique coordinate vector $\hat{\alphab} \in \F^n$ such that  $\hat{g} = \sum_i \hat{\alpha}_i \hat{b}_i$.  If  $g = \sum_j \alpha_j b_j$ and $\hat{g} = \sum_i \hat{\alpha}_i \hat{b}_i$, then 
  $$
  \inner{\hat{g}}{g} = \sum_i \sum_j \hat{\alpha}_i \alpha_j \inner{\hat{b}_i}{b_j} = \sum_i \hat{\alpha}_i \alpha_i = \hat{\alphab} \cdot \alphab;
  $$
  \ie the inner product $ \inner{\hat{g}}{g}$ is the dot product $\hat{\alphab} \cdot \alphab$ of the coordinate vectors.  In other words, dual bases define adjoint isomorphisms $G \lra \F^n$ and $\hat{G} \lra \F^n$ that preserve inner products.
  
As with groups, 
if $H$ is a subspace of $G$, then the orthogonal subspace $H^\perp$  is a subspace of $\hat{G}$, and the orthogonal subspace  to $H^\perp$ is $H$.  Moreover, $\dim H + \dim H^\perp = \dim G = \dim \hat{G}$.

     If $\G = \Pi_k G_k$ is a finite direct product of a collection of groups or vector spaces $G_k$, then the dual group or space to $\G$ is the finite direct product $\hat{\G} = \Pi_k \hat{G}_k$, and the pairing or inner product between $\gb \in \G$ and $\hat{\gb} \in \hat{\G}$ is given by the componentwise sum
  $$
  \inner{\hat{\gb}}{\gb} = \sum_k \inner{\hat{\gb}_k}{\gb_k}.
  $$
 If $\HH = \Pi_k H_k$ is a direct product of subgroups or subspaces $H_k \subseteq G_k$, then the orthogonal space is the direct product $\HH^\perp = \Pi_k H_k^\perp \subseteq \hat{\G}$. 
 
\subsection{Projection/cross-section duality}

Projection/cross-section duality is one of the most fundamental and useful duality relationships for linear and group codes \cite{F01}.  

Given a linear or group code $\CC \subseteq \A = \prod_{k \in \I_\A} \A_k$, let $\J \subseteq \I_\A$ be any subset of the symbol index set, and $\K = \I_\A - \J$ its complement.  Then $\A$ may be written as the direct product $\A^\J \times \A^{\K}$, where $\A^\J = \prod_{k \in \J} \A_k$ and $\A^{\K} = \prod_{k \in \K} \A_k$.  Correspondingly we may write any $\ab \in \A$ as a pair $(\ab^\J, \ab^{\K})$, where $\ab^\J \in \A^\J$ and $\ab^{\K} \in \A^{\K}$ are the \emph{projections} of $\ab$ on $\J$ and $\K$, respectively.

The \emph{projection map} $P_\J: \A \to \A^\J$ defined by $\ab \mapsto \ab^\J$ is evidently a homomorphism.  Given a linear or group code $\CC \subseteq \A$, the \emph{projection}  of $\CC$ onto $\A^\J$ is $\CC_{|\A^\J} = P_\J(\CC)$, a subgroup of $\A^\J$.  

The \emph{cross-section} $\CC_{:\A^\J}$ of $\CC$ on $\A^\J$ is defined as $\CC_{:\A^\J} = \{\ab^\J \in \A^\J \mid (\ab^\J, \zerob^{\K}) \in \CC\}$, which is evidently a subgroup of $\CC_{|\A^\J}$.  We note that $\CC_{:\A^\J}$ is trivially isomorphic to the kernel of the projection map $P_{\K}: \CC \to \A^{\K}$, a subcode of $\CC$;  therefore the following duality relationship is sometimes called projection/subcode duality.

\vspace{1ex}
\noindent
\textbf{Lemma} (\textbf{Projection/cross-section duality}).  If $\CC \subseteq \A^\J \times \A^\K$ and $\CC^\perp \subseteq \hat{\A}^\J \times \hat{\A}^\K$ are orthogonal linear or group block codes, then $\CC_{:\A^\J}$ and $(\CC^\perp)_{|\hat{\A}^\J}$ are orthogonal linear or group codes.

\vspace{1ex}
\noindent
\textit{Proof}.  Because pairings or inner products are componentwise sums, we have $\inner{\hat{\ab}}{\ab} = \inner{\hat{\ab}^\J}{\ab^\J} + \inner{\hat{\ab}^{\K}}{\ab^{\K}}$.  Thus  $(\hat{\ab}^\J, \hat{\ab}^\K) \perp (\ab^\J, \zerob^\K) \Leftrightarrow \hat{\ab}^\J \perp \ab^\J$.  We therefore have the following logical chain:
$$
\ab^\J \in \CC_{:\A^\J} \Leftrightarrow (\ab^\J, \zerob^{\K}) \in \CC  \Leftrightarrow (\ab^\J, \zerob^{\K}) \perp \CC^\perp \Leftrightarrow \ab^\J \perp (\CC^\perp)_{|\A^\J}.  \qquad \qed
$$

  \subsection{Dual realizations}\label{Section 2.9}

The dual realization of a normal linear or group realization is defined below.
The normal graph duality theorem of \cite{F01} states that the dual realization realizes the orthogonal code $\CC^\perp$ to $\CC$.
By now there exist several proofs of this theorem, the most elegant of which involve Fourier transforms of code indicator functions \cite{MK05, AM11, F11b}.  We give here a new elementary proof (similar to that of Koetter \cite{K02}), partly in order to introduce concepts and notation that will be useful later in Section \ref{Section 4}.

Consider the set $\Pi_i \CC_i$ of all possible combinations of codewords of the constraint codes $\CC_i$.  With a normal realization, every external variable $A_k$ is involved in precisely one constraint code, and every internal variable $S_j$ is involved in precisely two constraint codes.  Let us replace $S_j$ by a replica variable $S'_j$ in one of its two appearances (it does not matter which), and add an equality constraint $s'_j = s_j$.  Then each element of $\Pi_i \CC_i$ may be uniquely identified by a triple $(\ab, \sb, \sb') \in \A \times \SSS \times \SSS$.

A valid configuration $(\ab, \sb, \sb') \in \Pi_i \CC_i$ is one for which $\sb' = \sb$.  Thus the behavior of the realization is $\Bf = \{(\ab, \sb) \in \A \times \SSS \mid (\ab, \sb, \sb) \in \Pi_i \CC_i\}$.  This is illustrated in Fig.\ \ref{NGDT}(a).  

The \emph{dual realization} is defined as the realization with the same graph topology in which the external alphabets $\A_k$ are replaced by their duals $\hat{\A_k}$, the internal alphabets $\SSS_j$ are replaced by their duals $\hat{\SSS_j}$, the constraint codes $\CC_i$ are replaced by their orthogonal codes $\CC_i^\perp$, and each equality constraint $s_j' = s_j$ is replaced by the sign inversion constraint $\hat{s}_j' = -\hat{s}_j$, as illustrated  in Fig.\ \ref{NGDT}(b).  Thus the dual behavior may be written as $\Bf^\circ = \{(\hat{\ab}, \hat{\sb}) \in \hat{\A} \times \hat{\SSS} \mid (\hat{\ab}, \hat{\sb}, -\hat{\sb}) \in  \Pi_i \CC_i^\perp\}$.  Note the slight sign asymmetry between the definitions of the primal and dual behaviors.

Alternatively, a dual way of defining the primal behavior is as follows.  Following Koetter \cite{K02}, we may write $\Bf = \{(\ab, \sb) \in \A \times \SSS \mid (\ab, \sb, \sb) \perp \Pi_i \CC_i^\perp\}$;  that is, $\Bf$ is the set of all configurations $(\ab, \sb)$ such that $(\ab, \sb, \sb) \perp (\hat{\ab}, \hat{\sb}, \hat{\sb}')$ for all $(\hat{\ab}, \hat{\sb}, \hat{\sb}') \in \Pi_i \CC_i^\perp$, or equivalently $(\ab, \sb) \perp (\hat{\ab}, \hat{\sb} + \hat{\sb}')$.  Therefore if we define $\Bf^\perp$ as the image of  the homomorphism $\Sigma:  \prod_i {\CC}_i^\perp \to \hat{\A} \times \hat{\SSS}$ that is defined by the sum map $(\hat{\ab}, \hat{\sb}, \hat{\sb}') \mapsto (\hat{\ab}, \hat{\sb} + \hat{\sb}')$, then $\Bf$ is the orthogonal code to $\Bf^\perp$.

Notice now that $(\hat{\ab}, \hat{\sb}, \hat{\sb}') \in \prod_i \CC_i^\perp$ corresponds to a valid configuration $(\hat{\ab}, \hat{\sb}, -\hat{\sb})$ in the dual behavior if and only if it is mapped  by $\Sigma$ to $(\hat{\ab}, \zerob)$.  Therefore the code $\D$ realized by the dual behavior is precisely the cross-section $(\Bf^\perp)_{:\hat{\A}} = \{\hat{\ab} \in \hat{\A} \mid (\hat{\ab}, \zerob) \in \Bf^\perp\}$.  But by projection/cross-section duality, the orthogonal code to $\D$ must be the projection $\Bf_{|\A}$, which is  $\CC$;  \ie $\D = \CC^\perp$. 

 In summary, we have proved:

\vspace{1ex}
\noindent
\textbf{Lemma (Normal realization duality)}.  If a normal realization as in Fig.\ \ref{NGDT}(a) realizes a linear or group code $\CC$, then  its dual realization as in Fig.\ \ref{NGDT}(b) realizes the orthogonal code $\CC^\perp$.  \qed  \vspace{-1ex}

\section{Trimness and properness}

In this section we discuss trimness and properness of constraint codes.  
These are dual properties;  \ie $\CC_i$ is trim if and only if $\CC_i^\perp$ is proper.
We show that a realization is locally reducible if it has a constraint code that is not both trim and proper.
Finally, we show that a linear finite cycle-free realization is minimal if and only if all constraint codes are trim and proper.

\subsection{Trim-proper duality}

A constraint code $\CC_i$ will be called \emph{trim} if the projection of $\CC_i$ onto every state space $\SSS_j$ that is involved in $\CC_i$ is $\SSS_j$;  \ie if every such projection is surjective.  

A  constraint code $\CC_i$ will be called \emph{proper} if the value of any state variable $S_j$ involved in $\CC_i$ is determined by the values of all other variables;  \ie if for any set of values of all variables involved in $\CC_i$ other than $S_j$, at most one codeword in $\CC_i$ has those values.  In the linear or group case, $\CC_i$ is proper if and only if there is no codeword of $\CC_i$ whose support is a single state variable $S_j$;  \ie for all $\SSS_j$ the cross-section $(\CC_i)_{:\SSS_j}$ is trivial. 

This definition of ``proper" generalizes the traditional definition for a trellis realization, which is called proper \cite{M88}  (or ``biproper" \cite{V98}) if the symbol values associated with all transitions to or from a given state are all different.\footnote{In system theory, for conventional state-space (trellis) realizations, properness is sometimes called ``instantaneous observability," because then and only then is the next state determined by the current state and symbol.}  

As in \cite{GLW11b}, we observe that trimness and properness are dual properties:

\vspace{1ex}
\noindent
\textbf{Theorem 1} (\textbf{Trim-proper duality}).  A linear or group constraint code $\CC_i$ is trim if and only if  its orthogonal code $\CC_i^\perp$ is proper. 

\vspace{1ex}
\noindent
\textit{Proof}:  By projection/cross-section duality, the cross-section $(\CC_i^\perp)_{:\hat{\SSS}_j}$ is the orthogonal code to the projection $(\CC_i)_{|\SSS_j}$.  Thus $\{0\} \subset (\CC_i^\perp)_{:\hat{\SSS}_j}$ if and only if $(\CC_i)_{|\SSS_j} \subset \SSS_j$.\footnote{By ``$A \subset B$," we mean that $A$ is a proper subset of $B$.}  \qed \vspace{1ex}

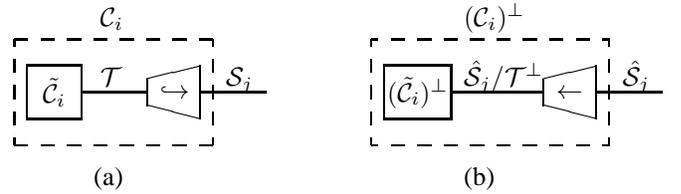
\begin{figure}[h]
\setlength{\unitlength}{5pt}
\centering
\begin{picture}(44,12)(-4, -5)
\put(-7,-3){\dashbox(15,8)}
\put(-0.5,6){$\CC_i$}
\put(-6,-1){\framebox(4,4){$\tilde{\CC}_i$}}
\put(-2,1){\line(1,0){5}}
\put(-0.5,1.5){$\T$}
\put(7,-1){\line(0,1){4}}
\put(4,0.5){$\hookrightarrow$}
\put(3,0){\line(0,1){2}}
\put(3,2){\line(4,1){4}}
\put(3,0){\line(4,-1){4}}
\put(-1,-6){(a)}
\put(7,1){\line(1,0){5}}
\put(9,1.5){$\SSS_j$}
\put(20,-3){\dashbox(18,8)}
\put(27,6){$(\CC_i)^\perp$}
\put(21,-1){\framebox(5,4){$(\tilde{\CC}_i)^\perp$}}
\put(26,1){\line(1,0){7}}
\put(27,1.5){$\hat{\SSS}_j/\T^\perp$}
\put(37,-1){\line(0,1){4}}
\put(34,0.5){$\leftarrow$}
\put(33,0){\line(0,1){2}}
\put(33,2){\line(4,1){4}}
\put(33,0){\line(4,-1){4}}
\put(27,-6){(b)}
\put(37,1){\line(1,0){5}}
\put(39,1.5){$\hat{\SSS}_j$}
\end{picture}
\caption{Dual realizations:  (a) nontrim  $\CC_i$;  (b) improper dual $(\CC_i)^\perp$.}
\label{TPD}
\end{figure}

This duality is illustrated in Fig.\ \ref{TPD}.  $\CC_i$ is not trim at $\SSS_j$ if and only if the projection $(\CC_i)_{|\SSS_j}$ is a proper subspace or subgroup $\T \subset \SSS_j$.  In this case $\CC_i$ may be represented as the concatenation of a trimmed constraint code $\tilde{\CC}_i$, in which $\SSS_j$ is replaced by $\T$ without changing any codewords, and an \emph{inclusion constraint code} $\CC_{\hookrightarrow} = \{(t, t) \in \T \times  \SSS_j: t \in \T\}$, as illustrated in Fig.\ \ref{TPD}(a).

By definition, the dual realization of the improper  code $(\CC_i)^\perp$ is obtained by concatenating the orthogonal code $(\tilde{\CC}_i)^\perp$, a sign inverter, and the orthogonal code $(\CC_{\hookrightarrow})^\perp$.  It is easy to see that  $(\CC_{\hookrightarrow})^\perp = \{(-\hat{s} + \T^\perp, \hat{s}) \in \hat{\SSS_j}/\T^\perp  \times \hat{\SSS_j}:  \hat{s} \in \hat{\SSS}_j\}$, using the inner product $\inner{\hat{s} + \T^\perp}{t} = \inner{\hat{s}}{t}$ for $t \in \T$ and $\hat{s} \in \hat{\SSS}_j$.\footnote{In other words, the inclusion map $\T_j \hookrightarrow \SSS_j$ and the natural map $\hat{\SSS}_j \rightarrow \hat{\SSS}_j/(\T_j)^\perp$ are adjoint homomorphisms.}  
Combining the sign inverter and $(\CC_{\hookrightarrow})^\perp$, we obtain the \emph{quotient constraint code} $\CC_{\leftarrow} = \{(\hat{s} + \T^\perp, \hat{s}) \in \hat{\SSS_j}/\T^\perp  \times \hat{\SSS_j}:  \hat{s} \in \hat{\SSS}_j\}$, which involves the natural map from $\hat{\SSS}_j$ to its quotient $\hat{\SSS}_j/\T^\perp$, as shown in Fig.\ \ref{TPD}(b). 

In view of Theorem 1, every improper constraint code may be represented as in Fig.\ \ref{TPD}(b) as a code with a reduced state space $\hat{\SSS}'_j = \hat{\SSS}_j/\T^\perp$, obtained by merging all states in each coset $\hat{s} + \T^\perp$, plus a constraint code which allows every coset  of $\T^\perp$ to branch to all of its elements.

\subsection{Local irreducibility requires trimness and properness}

We now show that both ``trim" and ``proper" are necessary for local irreducibility.

\vspace{1ex}
\noindent
\textbf{Theorem 2} (\textbf{Local reducibility}).  A normal  linear or group realization is locally reducible if any constraint code $\CC_i$ is not both trim and proper. 

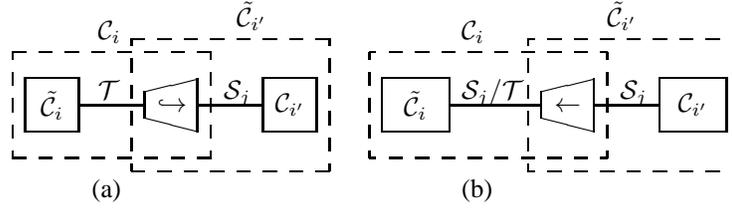
\begin{figure*}[t]
\setlength{\unitlength}{5pt}
\centering
\begin{picture}(44,12)(-3, -5)
\put(-7,-3){\dashbox(15,8)}
\put(-0.5,6){$\CC_i$}
\put(-6,-1){\framebox(4,4){$\tilde{\CC}_i$}}
\put(-2,1){\line(1,0){5}}
\put(-0.5,1.5){$\T$}
\put(2,-4){\dashbox(15,10)}
\put(7,-1){\line(0,1){4}}
\put(4,0.5){$\hookrightarrow$}
\put(3,0){\line(0,1){2}}
\put(3,2){\line(4,1){4}}
\put(3,0){\line(4,-1){4}}
\put(-1,-6){(a)}
\put(7,1){\line(1,0){5}}
\put(9,1.5){$\SSS_j$}
\put(10,7){$\tilde{\CC}_{i'}$}
\put(12,-1){\framebox(4,4){${\CC}_{i'}$}}
\put(20,-3){\dashbox(18,8)}
\put(27,6){$\CC_i$}
\put(21,-1){\framebox(5,4){$\tilde{\CC}_i$}}
\put(26,1){\line(1,0){7}}
\put(27,1.5){$\SSS_j/\T$}
\put(32,-4){\dashbox(16,10)}
\put(37,-1){\line(0,1){4}}
\put(34,0.5){$\leftarrow$}
\put(33,0){\line(0,1){2}}
\put(33,2){\line(4,1){4}}
\put(33,0){\line(4,-1){4}}
\put(27,-6){(b)}
\put(37,1){\line(1,0){5}}
\put(39,1.5){$\SSS_j$}
\put(38,7){$\tilde{\CC}_{i'}$}
\put(42,-1){\framebox(5,4){${\CC}_{i'}$}}
\end{picture}
\caption{Dual reductions:  (a) nontrim  $\CC_i$;  (b) improper $\CC_i$.}
\label{TPDa}
\end{figure*}

\vspace{1ex}
\noindent
\textit{Proof}:  
We give a pictorial proof, following Fig.\ \ref{TPD}.  Suppose $\CC_i$ is not trim at $\SSS_j$;  \ie $\T = (\CC_i)_{|\SSS_j} \subset \SSS_j$.  Let $\CC_{i'}$ be the other constraint in which $\SSS_j$ is involved.  Then, as shown in Fig.\ \ref{TPDa}(a), the combination of $\CC_i$ with $\CC_{i'}$ is equivalent to the combination of the trimmed code $\tilde{\CC}_i$ with a second trimmed code $\tilde{\CC}_{i'}$, namely the combination of $\CC_{i'}$ with $\CC_{\hookrightarrow}$, which restricts $\SSS_j$ to $\T$.  Replacing the combination of $\CC_i$ and $\CC_{i'}$ with that of $\tilde{\CC}_i$ and $\tilde{\CC}_{i'}$ thus strictly reduces the connecting state space from $\SSS_j$ to $\T$, without changing the code $\CC$ realized by the realization.  

Dually, suppose $\CC_i$ is not proper at $\SSS_j$;  \ie $\{0\} \subset \T = (\CC_i)_{:\SSS_j}$.  Again, let  $\CC_{i'}$ be the other constraint in which $\SSS_j$ is involved.  Then, as shown in Fig.\ \ref{TPDa}(b), the combination of $\CC_i$ with $\CC_{i'}$ is equivalent to the combination of the merged code $\tilde{\CC}_i$ with a second merged code $\tilde{\CC}_{i'}$, namely the combination of $\CC_{i'}$ with $\CC_{\leftarrow}$, which merges $\SSS_j$ to $\SSS_j/\T$ via the natural map.  Replacing the combination of $\CC_i$ and $\CC_{i'}$ with that of $\tilde{\CC}_i$ and $\tilde{\CC}_{i'}$ thus strictly reduces the connecting state space from $\SSS_j$ to $\SSS_j/\T$, without changing the code $\CC$ realized by the realization.  \qed \vspace{1ex}

Theorem 2 and its proof show that in any normal linear or group realization, if any constraint code is not trim or improper, then there is a strict reduction of the corresponding state space $\SSS_j$ via trimming or merging, respectively, such that only the two constraint codes in which $\SSS_j$ is involved are affected.  It is therefore hard to imagine any application in which one would not begin by ensuring that all constraint codes have been made trim and proper.

This development shows that  ``merging" (quotient-taking) is the dual reduction to ``trimming" (restricting).  Quotient-taking  preserves the linear or group property of the realization, and thus is the appropriate notion of ``merging" for linear or group realizations.  

For conventional and tail-biting trellis realizations, the fact that ``improper" implies ``mergeable" is well known;  see, \eg \cite{KV03}.

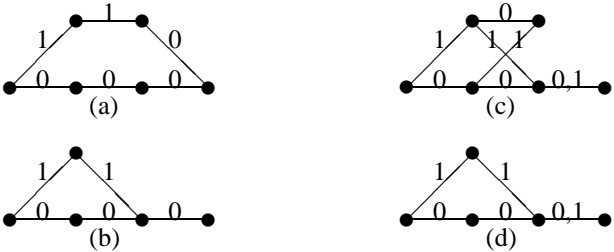
\begin{figure}[h]
\setlength{\unitlength}{5pt}
\centering
\begin{picture}(50,16)(-2, -1)
\put(0,10){\circle*{1}}
\put(0,10){\line(1,0){5}}
\put(0,10){\line(1,1){5}}
\put(2,10){0}
\put(2,13){1}
\put(5,10){\circle*{1}}
\put(5,15){\circle*{1}}
\put(5,15){\line(1,0){5}}
\put(5,10){\line(1,0){5}}
\put(7,10){0}
\put(7,15){1}
\put(10,10){\circle*{1}}
\put(10,15){\circle*{1}}
\put(10,15){\line(1,-1){5}}
\put(10,10){\line(1,0){5}}
\put(12,10){0}
\put(12,13){0}
\put(15,10){\circle*{1}}
\put(6,8){(a)}
\put(0,0){\circle*{1}}
\put(0,0){\line(1,0){5}}
\put(0,0){\line(1,1){5}}
\put(2,0){0}
\put(2,3){1}
\put(5,0){\circle*{1}}
\put(5,5){\circle*{1}}
\put(5,5){\line(1,-1){5}}
\put(5,0){\line(1,0){5}}
\put(7,0){0}
\put(7,3){1}
\put(10,0){\circle*{1}}
\put(10,0){\line(1,0){5}}
\put(12,0){0}
\put(15,0){\circle*{1}}
\put(6,-2){(b)}
\put(30,10){\circle*{1}}
\put(30,10){\line(1,0){5}}
\put(30,10){\line(1,1){5}}
\put(32,10){0}
\put(32,13){1}
\put(35,10){\circle*{1}}
\put(35,15){\circle*{1}}
\put(35,15){\line(1,0){5}}
\put(35,10){\line(1,0){5}}
\put(35,15){\line(1,-1){5}}
\put(35,10){\line(1,1){5}}
\put(37,10){0}
\put(37,15){0}
\put(36,13){1}
\put(38,13){1}
\put(40,10){\circle*{1}}
\put(40,15){\circle*{1}}
\put(40,10){\line(1,0){5}}
\put(41,10){0,1}
\put(45,10){\circle*{1}}
\put(36,8){(c)}
\put(30,0){\circle*{1}}
\put(30,0){\line(1,0){5}}
\put(30,0){\line(1,1){5}}
\put(32,0){0}
\put(32,3){1}
\put(35,0){\circle*{1}}
\put(35,5){\circle*{1}}
\put(35,5){\line(1,-1){5}}
\put(35,0){\line(1,0){5}}
\put(37,0){0}
\put(37,3){1}
\put(40,0){\circle*{1}}
\put(40,0){\line(1,0){5}}
\put(41,0){0,1}
\put(45,0){\circle*{1}}
\put(36,-2){(d)}
\end{picture}
\caption{(a) Improper, (b) merged, (c) non-trim and (d) trimmed realizations of dual codes.}
\label{Fig1a}
\end{figure}

\vspace{1ex}
\noindent
\textbf{Example 1} (compare \cite[Fig.\ 2]{V98}).  Consider the binary linear block code $\CC = \{000, 110\}$.  This code may be realized by the linear trellis realization shown in Fig.\ \ref{Fig1a}(a), with three binary symbol alphabets, four state spaces $\SSS_0 = \SSS_3 = \{0\}, \SSS_1 = \SSS_2 = \{0,1\}$, and three constraint codes $\CC_0 = \{000, 011\}, \CC_1 = \{000, 111\}$, and $\CC_2 = \{000, 100\}$.  Since $\CC_2$ has a nonzero codeword 100 supported by the single state space $\SSS_2$, $\CC_2$ is improper, and therefore can be reduced.  Indeed, the two states of $\SSS_2$ can be merged to a single state $\{0\}$ without affecting $\CC$;  this yields the minimal linear trellis realization of this code, shown in Fig.\ \ref{Fig1a}(b).

The dual linear trellis realization to that of Fig.\ \ref{Fig1a}(a) uses the same variable alphabets and the orthogonal constraint codes $\CC_0^\perp = \{000, 011\}, \CC_1^\perp = \{000, 011, 110, 101\}$, and $\CC_2^\perp = \{000, 010\}$, as shown in Fig.\ \ref{Fig1a}(c). (Because the ground field is binary, no sign inverters are needed.)  It can be seen that this dual realization, although unusual, does indeed realize $\CC^\perp = \{000, 110, 001,$ $111\}$.  Since the projection of $\CC_2^\perp$ onto $\hat{\SSS}_2$ is $\{0\}$, $\CC_2^\perp$ is not trim, and therefore can be reduced.    Indeed, restricting $\hat{\SSS}_2$ to $\{0\}$, we obtain the minimal trellis realization of $\CC^\perp$ of Fig.\ \ref{Fig1a}(d).  \qed 

\subsection{Minimal cycle-free realizations}\label{Section 3.3}

It is well known that a finite conventional linear trellis realization is minimal if and only if  all end state spaces are trivial and every constraint code $\CC_i$ is trim and proper  \cite{V98}.  We now show that  this result holds for any finite cycle-free linear or group realization.  As a byproduct, we obtain a constructive proof of the State Space Theorem, which is perhaps the most fundamental result of linear behavioral system theory \cite{W89}.

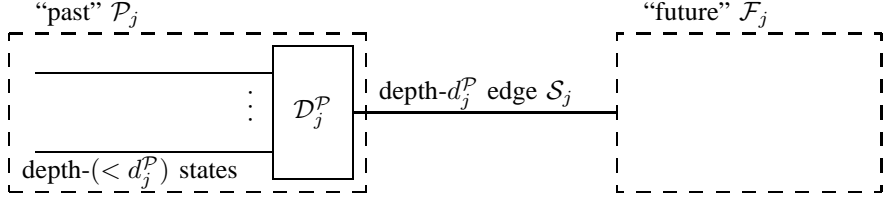
\begin{figure*}[t]
\setlength{\unitlength}{5pt}
\centering
\begin{picture}(40,14)(0, -1)
\put(-10,2){\line(1,0){18}}
\put(-10,8){\line(1,0){18}}
\put(6,4){$\cdot$}
\put(6,5){$\cdot$}
\put(6,6){$\cdot$}
\put(8,0){\framebox(6,10){$\D_j^\PP$}}
\put(-11,0){depth-$(< d_j^\PP)$ states}
\put(14,5){\line(1,0){20}}
\put(16,6){depth-$d_j^\PP$ edge $\SSS_j$}
\put(-12,-1){\dashbox(27,12){}}
\put(-10,12){``past" $\PP_j$}
\put(34,-1){\dashbox(20,12){}}
\put(36,12){``future" $\FF_j$}
\end{picture}
\caption{Illustration of the past $\PP_j$ and future $\FF_j$ of an edge $\SSS_j$.}
\label{MCF}
\end{figure*}

The key graph-theoretic property of cycle-free graphs is that every edge is a cut set.  In other words, if $\SSS_j$ is any state space in a normal cycle-free realization, then cutting the edge corresponding to $\SSS_j$ partitions the normal graph of the realization into two disconnected subgraphs, which we label arbitrarily as $\PP_j$ and $\FF_j$ (for ``past" and ``future"), as shown in Fig.\ \ref{MCF}.

Furthermore, if we define the \emph{past depth} $d_j^\PP$ of the edge $\SSS_j$ as the maximum distance to any leaf node in $\PP_j$, then the depths of all other states involved in the constraint code $\D_j^\PP$ in $\PP_j$ to which $\SSS_j$ is connected must be less than $d_j^\PP$  (see Fig.\ \ref{MCF}).  (Again, we assume there are no degree-1 states.)  

We correspondingly represent the symbol configuration space $\A$ as the product $\A^{\PP_j} \times \A^{\FF_j}$ of past and future components, and write a symbol configuration as a pair $\ab = (\ab^{\PP_j}, \ab^{\FF_j})$.  The projection and cross-section of the code $\CC$ on the past are written as $\CC_{|\A^{\PP_j}}$ and $\CC_{:\A^{\PP_j}}$, respectively.  We similarly define past and future state configurations $\sb^{\PP_j} \in \SSS^{\PP_j}$ and  $\sb^{\FF_j} \in \SSS^{\FF_j}$, where $\SSS_j$ is included in both $\SSS^{\PP_j}$ and $\SSS^{\FF_j}$.  Finally, we 
say that a past configuration $\ab^{\PP_j}$ \emph{reaches} a state $s_j \in \SSS_j$ if there is a past configuration $(\ab^{\PP_j}, \sb^{\PP_j})$ that satisfies all past constraints and includes $s_j$.

We then have the following remarkably simple result:

\vspace{1ex}
\noindent
\textbf{Theorem 3} (\textbf{Minimal = trim + proper}).
Given a normal linear or group realization of a code $\CC$ on a finite connected cycle-free graph $\G$, the following are equivalent:
\begin{itemize}
\item[(1)]   The realization is minimal.
\item[(2)]  Every constraint code $\CC_i$ is both trim and proper.
\item[(3)]  Every state space $\SSS_j$ is isomorphic to the ``past-induced state space" $\CC_{|\A^{\PP_j}}/\CC_{:\A^{\PP_j}}$, and the set of $\ab^{\PP_j}$ that reach a given state $s_j \in \SSS_j$ is the corresponding coset of $\CC_{:\A^{\PP_j}}$ in $\CC_{|\A^{\PP_j}}$.
\end{itemize}

\pagebreak

\vspace{1ex}
\noindent
\textit{Proof}:  

(1 $\Rightarrow$ 2)  By Theorem 2, if any $\CC_i$ is not both trim and proper, then the realization is locally reducible, and thus not minimal.   

(2 $\Rightarrow$ 3)  Let $(\CC^{\PP_j})_{|\A^{\PP_j}}$ denote the set of all past configurations $\ab^{\PP_j} \in \A^{\PP_j}$ that appear in some valid past configuration $(\ab^{\PP_j}, \sb^{\PP_j})$.  We prove by finite induction on $d_j^\PP$ that
\begin{enumerate}
\item[(a)] trim $\Rightarrow$ every state $s_j \in \SSS_j$ is reached by some symbol configuration $\ab^{\PP_j} \in (\CC^{\PP_j})_{|\A^{\PP_j}}$; 
\item[(b)] proper $\Rightarrow$ every symbol configuration $\ab^{\PP_j} \in (\CC^{\PP_j})_{|\A^{\PP_j}}$ reaches a unique state $s_j \in \SSS_j$.  
\end{enumerate}
In other words, a trim and proper realization realizes a well-defined surjective ``reaching map" $R: (\CC^{\PP_j})_{|\A^{\PP_j}} \to \SSS_j$.

Clearly (a) and (b) hold if the depth of $\SSS_j$ is 1;  \ie if $\PP_j$ comprises a single leaf node, representing a constraint that involves no state other than $\SSS_j$.  Now if (a) and (b) hold for all depth-($< d_j^\PP$) edges in $\PP_j$, then (a) and (b) hold for all depth-$d_j^\PP$ edges $\SSS_j$, because, denoting the constraint code to which $\SSS_j$ is connected in $\PP_j$ as $\D_j^\PP$   (see Fig.\ \ref{MCF}):  
\begin{enumerate}
\item[(a)] if $\D_j^\PP$ is trim, and all of its depth-$(<d_j^\PP)$ states are reachable, then all states $s_j \in \SSS_j$ are reachable, since $s_j$ occurs with some configuration of depth-$(<d_j^\PP)$ state values in some codeword of $\D_j^\PP$;
\item[(b)] if $\D_j^\PP$ is proper, and every symbol configuration in the ``past" of its depth-$(<d_j^\PP)$ states reaches a unique state value, then the same must be true for $\SSS_j$, since otherwise there would be two codewords of $\D_j^\PP$ that differ only on $\SSS_j$.
\end{enumerate}
In particular, the all-zero symbol configuration $\zerob^{\PP_j}$ reaches only the zero state of $\SSS_j$.

Now, applying the same argument to $\FF_j$, we conclude that all states of $\SSS_j$ are reached by future configurations, so all valid past configurations are past projections of codewords;  \ie $(\CC^{\PP_j})_{|\A^{\PP_j}} = \CC_{|\A^{\PP_j}}$. Moreover, the unique state of $\SSS_j$ that can be reached by the all-zero future symbol configuration $\zerob^{\FF_j}$ is the zero state.  Hence the set of configurations in $\CC_{|\A^{\PP_j}}$ that reach the zero state in $\SSS_j$ is precisely the cross-section $\CC_{:\A^{\PP_j}} = \{\ab^{\PP_j} \in \CC_{|\A^{\PP_j}} \mid (\ab^{\PP_j}, \zerob^{\FF_j}) \in \CC\}$.  

By linearity, the ``reaching map" $R: \CC_{|\A^{\PP_j}} \to \SSS_j$ is a homomorphism.  We have shown that it is surjective and has kernel $\CC_{:\A^{\PP_j}}$.  Thus, by the fundamental theorem of homomorphisms, $\SSS_j \cong \CC_{|\A^{\PP_j}}/\CC_{:\A^{\PP_j}}$.  Moreover, since the subset of configurations in $\CC_{|\A^{\PP_j}}$ that reach the zero state in $\SSS_j$ is $\CC_{:\A^{\PP_j}}$, by linearity the subset of configurations in $\CC_{|\A^{\PP_j}}$ that reach an arbitrary state in $\SSS_j$ is the corresponding coset of $\CC_{:\A^{\PP_j}}$.  

(3 $\Rightarrow$ 1)  Since (3) is equally valid for $\FF_j$ and $\PP_j$, it follows that $\SSS_j$ is also isomorphic to the ``future-induced state space" $\CC_{|\A^{\FF_j}}/\CC_{:\A^{\FF_j}}$, and the set of $\ab^{\FF_j} \in \CC_{|\A^{\FF_j}}$ that reach a given state $s_j \in \SSS_j$ is the corresponding coset of $\CC_{:\A^{\FF_j}}$ in $\CC_{|\A^{\FF_j}}$.

Thus if $\{\rb^{\PP_j}(s_j): s_j \in \SSS_j\}$ is a set of coset representatives for the cosets of $\CC_{:\A^{\PP_j}}$ in $\CC_{|\A^{\PP_j}}$, and $\{\rb^{\FF_j}(s_j): s_j \in \SSS_j\}$ is a set of coset representatives for the cosets of $\CC_{:\A^{\FF_j}}$ in $\CC_{|\A^{\FF_j}}$, then the set of all symbol configurations $(\ab^{\PP_j}, \ab^{\FF_j}) \in \A^{\PP_j} \times \A^{\FF_j}$ that pass through a given state $s_j \in \SSS_j$ is $(\rb^{\PP_j}(s_j), \rb^{\FF_j}(s_j))  + \CC_{:\A^{\PP_j}} \times  \CC_{:\A^{\FF_j}}$, a coset of $\CC_{:\A^{\PP_j}}\times \CC_{:\A^{\FF_j}}$.  The code $\CC$ is thus the union of these $|\SSS_j|$ disjoint cosets.  (An easy corollary is that $\SSS_j$ is also isomorphic to $\CC/(\CC_{:\A^{\PP_j}}\times \CC_{:\A^{\FF_j}})$.)

This shows that every realization of~$\CC$ with the same graph topology must have at least $|\SSS_j|$
states at time~$j$, because $(\ab^{\PP_j}, \ab^{\FF_j}) \in \A^{\PP_j} \times \A^{\FF_j}$ is in $\CC$ if and only if $\ab^{\PP_j}$ and $\ab^{\FF_j}$ are in corresponding cosets of $\CC_{:\A^{\PP_j}}$ and $\CC_{:\A^{\FF_j}}$.  Thus the realization is minimal.  \qed 
\vspace{1ex} 

This proof shows constructively that in a minimal cycle-free realization there is a one-to-one match of the cosets of $\CC_{:\A^{\PP_j}}$ in $\CC_{|\A^{\PP_j}}$ with the cosets of $\CC_{:\A^{\FF_j}}$ in $\CC_{|\A^{\FF_j}}$ such that every $\ab^{\PP_j}$ in a given coset can be followed by every $\ab^{\FF_j}$ in its matching coset to form a codeword in $\CC$, and no other pairs $(\ab^{\PP_j}, \ab^{\FF_j})$ are codewords in $\CC$.  This is the essence of the State Space Theorem \cite{W89, FT93, F01}.  This theorem also yields the SST isomorphisms
$$
\frac{\CC_{|\A^{\PP_j}}}{\CC_{:\A^{\PP_j}}} \cong \frac{\CC_{|\A^{\FF_j}}}{\CC_{:\A^{\FF_j}}} \cong \frac{\CC}{\CC_{:\A^{\PP_j}} \times \CC_{:\A^{\FF_j}}},
$$
and the fact that a minimal state space $\SSS_j$ is isomorphic to any of these quotients.

In general, we will say that a realization is \emph{trim} (resp.\ \emph{proper}) if every constraint code is trim (resp.\ proper).  We will further define a realization to be \emph{state-trim} if the projection of the full behavior $\Bf$ onto every state space $\SSS_j$ is surjective (\ie equal to $\SSS_j$), and \emph{branch-trim} if the projection of $\Bf$ onto every constraint code $\CC_i$ is surjective.  Thus ``trim" and ``proper" are local properties, since they involve only local constraint codes, whereas ``state-trim" and ``branch-trim" are global, since they involve the full behavior.  
Notice that the second part of this proof shows that for a finite cycle-free realization, ``trim" implies ``state-trim" and ``branch-trim."

Koetter \cite[Lemma 6]{K02} shows that a state space $\SSS_j$ may be reduced by merging if the dual behavior $\Bf^\circ$ is not state-trim at $\hat{\SSS}_j$, a global condition that yields our local result ``improper" $\Rightarrow$ ``locally reducible by merging" as a corollary.

Kashyap \cite{K09a} has shown that a given tree realization may be minimized by first making it state-trim (``essential"), and then merging each state space $\SSS_j$ into its quotient $\SSS_j/W_j$, where $W_j$ is the subspace of $\SSS_j$ that is reached by configurations in the zero-state subcode $\CC_{:\A^{\PP_j}} \times \CC_{:\A^{\FF_j}}$.  However, because the latter step requires computing the full behavior,  this procedure is not  very efficient.

From Theorem 3, it follows that if a linear or group realization on a finite cycle-free graph is nonminimal, then it may be made minimal by a finite sequence of local reductions.  In other words,  there exists a straightforward and efficient finite algorithm for minimizing any given linear or group realization on any  finite cycle-free graph.

From Theorems 1 and 3, it follows that the dual to a finite cycle-free minimal realization is minimal.  The dual minimal state spaces are thus the dual groups or spaces to the primal minimal state spaces;  \ie we obtain the dual state space theorem \cite{FT04} as a corollary.

Finally, we remark that the ``shortest basis" approach to minimality that is often used for conventional linear trellis realizations (see \cite{F11} and references therein) cannot be extended to general cycle-free realizations, because it relies on the ``product factorization," which generally does not exist for general cycle-free graphs;  see Appendix A.  In this respect, the ``trim and proper" approach to minimality may be regarded as more basic than the ``shortest basis" approach.

\section{Observability and controllability}\label{Section 4}

In this section, we define a realization to be controllable if its constraints are independent, and  give a simple test for controllability.  We show that a realization is uncontrollable if and only if the dual realization is unobservable, and that in either case such a realization is locally reducible. 

  We give conditions for controllability and observability for finite cycle-free realizations,  generator realizations, parity-check realizations and  tail-biting trellis realizations.  A tail-biting trellis realization is uncontrollable if and only if its behavior consists of disconnected subbehaviors, as with classical uncontrollable conventional trellis realizations;  however, on more general graphs, uncontrollability does not necessarily imply disconnectedness.  We show that the support of an unobservable configuration must be a cycle or a generalized cycle. 
  
 We conclude by observing that for iterative decoding, unobservable realizations seem clearly undesirable;  however, uncontrollability may not hurt, and may even be advantageous.

\subsection{Independent constraints and controllability}

The behavior~$\Bf$ of a linear realization is defined by a system of linear homogeneous
constraint equations.  
Specifically, each constraint code~$\CC_i,\,i\in\I_{\CC}$, of the realization may be specified by $\dim\CC_i^{\perp}$ independent equations, so $\Bf$ is the solution space of $\sum_{i\in\I_{\CC}}\dim\CC_i^{\perp}$ homogeneous equations.
We say that the realization has \emph{independent constraints} if the system of all
these equations is linearly independent.

We then have the following fundamental duality theorem:

\vspace{1ex}
\noindent
\textbf{Theorem 4}  (\textbf{Observability/controllability duality}).
A normal linear or group realization has independent constraints if and only if its dual realization is observable.

\vspace{1ex}
\noindent
\textit{Proof}: 
Suppose the system of constraints is linearly dependent;  \ie there exists a nontrivial linear combination $\sum_{i}(\hat{\ab}^{(i)},\hat{\sb}^{(i)})=(\zerob,\zerob)$ with $(\hat{\ab}^{(i)},\hat{\sb}^{(i)}) \in (\CC_i)^\perp$ for all $i$.
Since each symbol variable is involved in exactly one constraint code,
this implies $\hat{a}_k=0$ for all~$k$.
Moreover, since each state variable is involved in exactly two constraint codes, this implies that the two corresponding values  $\hat{s}_j$ and $\hat{s}'_j$ satisfy $\hat{s}_j=-\hat{s}'_j$.
But all this implies that $(\zerob,\hat{\sb})$ is a valid configuration in the dual realization (see Fig.\ \ref{NGDT}(b)).  Thus the dual realization is unobservable.
The converse follows by reversing these arguments.
\qed \vspace{1ex}

For classical conventional state (trellis) realizations, the dual property to observability is called controllability.  We will therefore call a linear or group realization that has independent constraints \emph{controllable}.  We use this term even though (a) the classical definition arose at a time when linear system theory was embedded in control theory, which is not our context here;  (b) although the behavior of uncontrollable tail-biting trellis realizations is similar to that of uncontrollable conventional trellis realizations, as we will see in Theorem 10, such uncontrollability properties do not necessarily extend to realizations on general graphs, as we will see in Section \ref{Section 4.7}.  The reader who is not so interested in continuity with classical linear system theory may therefore prefer terms like ``one-to-one" and ``independent" to ``observable" and ``controllable."

\begin{figure}[h]
\setlength{\unitlength}{5pt}
\centering
\begin{picture}(50,7)(-2, 9)
\put(30,10){\circle*{1}}
\put(30,15){\circle*{1}}
\put(30,10){\line(1,0){5}}
\put(30,15){\line(1,0){5}}
\put(32,10){0}
\put(32,15){1}
\put(35,10){\circle*{1}}
\put(35,15){\circle*{1}}
\put(35,15){\line(1,0){5}}
\put(35,10){\line(1,0){5}}
\put(37,10){0}
\put(37,15){1}
\put(40,10){\circle*{1}}
\put(40,15){\circle*{1}}
\put(40,15){\line(1,0){5}}
\put(40,10){\line(1,0){5}}
\put(42,10){0}
\put(42,15){1}
\put(45,10){\circle*{1}}
\put(45,15){\circle*{1}}
\put(6,8){(a)}
\put(0,10){\circle*{1}}
\put(0,15){\circle*{1}}
\put(0,10){\line(1,0){5}}
\put(0,15){\line(1,0){5}}
\put(0,10){\line(1,1){5}}
\put(0,15){\line(1,-1){5}}
\put(2,10){0}
\put(2,15){0}
\put(1,13){1}
\put(3,13){1}
\put(5,10){\circle*{1}}
\put(5,15){\circle*{1}}
\put(5,15){\line(1,0){5}}
\put(5,10){\line(1,0){5}}
\put(5,15){\line(1,-1){5}}
\put(5,10){\line(1,1){5}}
\put(7,10){0}
\put(7,15){0}
\put(6,13){1}
\put(8,13){1}
\put(10,10){\circle*{1}}
\put(10,15){\circle*{1}}
\put(10,10){\line(1,0){5}}
\put(10,15){\line(1,0){5}}
\put(10,10){\line(1,1){5}}
\put(10,15){\line(1,-1){5}}
\put(12,10){0}
\put(12,15){0}
\put(11,13){1}
\put(13,13){1}
\put(15,10){\circle*{1}}
\put(15,15){\circle*{1}}
\put(36,8){(b)}
\end{picture}
\caption{(a) Unobservable and (b) uncontrollable tail-biting realizations of orthogonal codes.}
\label{Fig2a}
\end{figure}
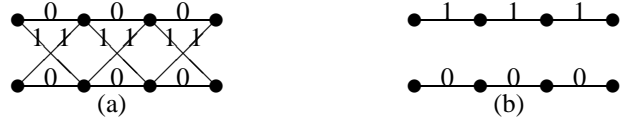

\vspace{1ex}
\noindent
\textbf{Example 2} (\cf \cite[Fig.\ 5]{KV03}).  The binary linear $(3,2,2)$ block code $\CC = \{000, 110, 101, 011\}$ may be realized by the linear tail-biting trellis realization shown in Fig.\ \ref{Fig2a}(a), with three binary symbol alphabets, three binary state spaces $\SSS_0 =  \SSS_1 = \SSS_2 = \{0,1\}$, and three constraint codes $\CC_0 = \CC_1$ $ = \CC_2 = \{000, 110, 101, 011\}$, where $\CC_2$ involves $\SSS_2$ and $\SSS_0$. Because the all-zero codeword $\zerob$ is realized by two configurations, this realization is unobservable, with $\dim \SSS^u = 1$.   

The dual linear tail-biting trellis realization to that of Fig.\ \ref{Fig2a}(a) uses the same variable alphabets and the orthogonal constraint codes $\CC_0^\perp = \CC_1^\perp = \CC_2^\perp = \{000, 111\}$, shown in Fig.\ \ref{Fig2a}(b). (Again, because the field is binary, no sign inverters are needed.)  The dual realization realizes the orthogonal $(3,1,3)$ code $\CC^\perp = \{000, 111\}$;  however, since it is the dual to an unobservable realization, it is uncontrollable by Theorem 4.   Explicitly, the three constraint equations  corresponding to the branches $101$ in the three primal constraint codes (\ie the constraints that form the unobservable configuration $(\ab, \sb) = (\zerob, \oneb)$ in the primal behavior) are dependent. \qed

\subsection{Finite cycle-free realizations}

For the finite cycle-free realizations that were considered in Section \ref{Section 3.3}, we have immediately:

\vspace{1ex}
\noindent
\textbf{Theorem 5}  (\textbf{Finite cycle-free realizations}). A normal linear or group realization on a finite cycle-free graph is observable if it is proper, and controllable if it is trim.  

\vspace{1ex}
\noindent
\textit{Proof}.  From the second part of the proof of Theorem 3, part (b), if a realization on a finite cycle-free graph is proper, then every symbol configuration $\ab \in \CC$ maps to a unique state configuration $\sb \in \SSS$, so the realization is one-to-one, and thus observable.
It then follows from the dualities of Theorems 1 and 4 that under the same conditions a trim realization  is controllable.  \qed \vspace{1ex}

Thus a minimal (trim and proper) finite cycle-free linear or group realization is observable and controllable.  However, the converse is not true:

\vspace{1ex}
\noindent
\textbf{Example 1} (cont.).  The trellis realization of Fig.\ \ref{Fig1a}(a) is observable, but improper.  The dual realization of Fig.\ \ref{Fig1a}(c) is controllable, but not trim.  \qed  
\vspace{1ex}

Thus in this context properness is stronger than observability, and trimness than controllability.  Properness implies invertibility not only globally, but also for fragments of a realization created by cuts.  Similarly, trimness implies state reachability not only globally, but also for fragments.

\subsection{Controllability test}

For a direct test of controllability, we regard $\prod_i {\CC}_i^\perp$ as a subspace of $\hat{\A} \times \hat{\SSS} \times \hat{\SSS}$, as in Fig.\ \ref{NGDT}(b).  We observe that the subspace $\K = \{(\zerob, \hat{\sb}, -\hat{\sb}) \in \prod_i {\CC}_i^\perp\}$ is isomorphic to the unobservable state configuration subspace $\hat{\SSS}^u = \{\hat{\sb} \in \hat{\SSS} \mid (\zerob, \hat{\sb}, -\hat{\sb}) \in \prod_i {\CC}_i^\perp\}$ of the dual realization.  Then we obtain:

\vspace{1ex}
\noindent
\textbf{Theorem 6}  (\textbf{Controllability test}).  The size of the behavior $\Bf$ of a linear or group realization with state configuration space $\SSS = \prod_j \SSS_j$ and constraint codes $\CC_i$ is $|\Bf| = ({\prod_i |\CC_i|})|\hat{\SSS}^u|/{|\SSS|}$;  or, in the linear case, $\dim \Bf = \sum_i \dim \CC_i + \dim \hat{\SSS}^u - \dim \SSS$.  Thus the realization is controllable if and only if $|\Bf| = (\prod_i |\CC_i|)/{|\SSS|} = (\prod_i |\CC_i|)/(\prod_j |\SSS_j|)$; or equivalently, in the linear case, if and only if 
$$\dim \Bf = \sum_i \dim \CC_i - \dim \SSS = \sum_i \dim \CC_i - \sum_j \dim \SSS_j.$$

\vspace{1ex}
\noindent
\textit{Proof}.  As in Section \ref{Section 2.9}, define $\Bf^\perp$ as the image of  the homomorphism $\Sigma:  \prod_i {\CC}_i^\perp \to \hat{\A} \times \hat{\SSS}$ such that $(\hat{\ab}, \hat{\sb}, \hat{\sb}') \mapsto (\hat{\ab}, \hat{\sb} + \hat{\sb}')$;  then $\Bf^\perp$ is the orthogonal code to $\Bf$.
Since $\K$ is the kernel of $\Sigma$, we have $\Bf^\perp \cong (\Pi_i \CC_i^\perp)/\K$ by the fundamental theorem of homomorphisms, so $|\Bf^\perp| = (\prod_i |\CC_i^\perp|)/|\K|$. 
Since $|\Bf| = |\A||\SSS|/|\Bf^\perp|$, $|\CC_i| = |\A^{(i)}||\SSS^{(i)}|/|\CC_i^\perp|$ and $|\K| = |\hat{\SSS}^u|$, and moreover, as a consequence of  the normal degree restrictions, $\prod_i |\A^{(i)}| = |\A|$ and $\prod_i |\SSS^{(i)}| = |\SSS|^2$, we have $|\Bf| = ({\prod_i |\CC_i|})|\hat{\SSS}^u|/{|\SSS|}$;  or, in the linear case, $\dim \Bf = \sum_i \dim \CC_i + \dim \hat{\SSS}^u - \dim \SSS$.  Thus $|\Bf| \ge ({\prod_i |\CC_i|})/{|\SSS|}$,
with equality if and only if $|\hat{\SSS}^u| = 1$;  \ie if and only if the realization is controllable.  \qed 

\vspace{1ex}
\noindent
\textbf{Example 2} (cont.).  In Fig.\ \ref{Fig2a}(b), we have $\dim \Bf^\circ = 1$, $\sum_i \dim \CC_i^\perp = 3$ and $\sum_i \dim \hat{\SSS}_i = 3$;  therefore by Theorem 6 this realization is uncontrollable.  However, for Fig.\ \ref{Fig2a}(a), we have $\dim \Bf =3$,  $\sum_i \dim \CC_i = 6$ and $\sum_i \dim \SSS_i = 3$, so this realization is controllable. \qed \vspace{1ex}

 If a linear realization is both observable and controllable, then we have $\dim \CC = \sum_i \dim \CC_i - \sum_j \dim \SSS_j$, since $\dim \Bf = \dim \CC$.   As  Kashyap \cite{K09b} observed, this implies that for minimal finite cycle-free linear realizations, the two complexity measures $\sum_i \dim \CC_i$ and $\sum_j \dim \SSS_j$ are equivalent.

\subsection{Generator and parity-check realizations}

We now investigate the observability and controllability of two standard types of realizations. 

A \emph{generator realization} of a linear code $\CC \subseteq \F^n$ is specified by a set of $\ell$ generator $n$-tuples $\gb_i \in \F^n$, such that $\CC$ is the set of all linear combinations $\ab = \sum_i \alpha_i \gb_i$ as the coefficient $\ell$-tuple $\alphab$ runs through $\F^\ell$.  The generators are \emph{linearly independent} if  $\dim \CC = \ell$.  A corresponding generator realization has (up to) $n$ internal replicas $\alpha_{ik}$ of each of the $\ell$ free coefficients $\alpha_i$, constrained to be equal by $\ell$ equality constraints $\alpha_{i1} = \cdots = \alpha_{in}$;  $n$ external variables $A_k$, and $n$ linear constraint codes that enforce the constraints $a_k = \sum_i \alpha_{ik} g_{ik} = \sum_i \alpha_i g_{ik}$.  (If $g_{ik} = 0$, then the replica $\alpha_{ik}$ may be omitted.)

A \emph{parity-check realization} of a linear code $\CC \subseteq \F^n$ is specified by a set of $r$ check $n$-tuples $\hb_j \in \F^n$, such that $\CC$ is set of all $n$-tuples $\ab \in \F^n$ that are orthogonal to all check $n$-tuples $\hb_j$.  In other words, $\CC$ is the orthogonal code to the linear code $\CC^\perp$ generated by the $r$ check $n$-tuples.  The checks are linearly independent if $\dim \CC^\perp = r$, or equivalently if $\dim \CC = n - r$.  In the corresponding parity-check realization, there are $n$ external variables $A_k$, $n$ one-dimensional constraint codes that generate (up to) $r$ multiples $a_k h_{jk}$ of each of the external variables $a_k$, and $r$ zero-sum constraint codes that enforce the parity checks $0 = \sum_k a_{k} h_{jk}$.  (If $h_{jk} = 0$, then the multiple $a_k h_{jk}$ may be omitted.)  

As expected, generator and parity-check realizations are duals:

\vspace{1ex} 
\noindent
\textbf{Theorem 7 (Generator/parity-check realization duality)}.  Let a generator realization of a linear code $\CC$ be specified by $\ell$ generator $n$-tuples $\gb_i$; then its dual realization is the parity-check realization specified by the same set of $\ell$ $n$-tuples as checks, which realizes the orthogonal code $\CC^\perp$.

\vspace{1ex}
\noindent
\textit{Proof}.  To prove that the realizations are duals, one need only check that:
\begin{itemize}
\item[(a)]  the orthogonal code to the code generated by the equality constraint $\alpha_{i1} = \cdots = \alpha_{in}$ is the code generated by the zero-sum constraint $\sum_k \hat{\alpha}_{ik} = 0$;
\item[(b)]  the constraint code defined by a check $a_k = \sum_i \alpha_{ik} g_{ik}$ on the 
variables $a_k, \alpha_{1k}, \ldots, \alpha_{\ell k}$ is the dual code to the code $\{(\hat{a}_k, -\hat{a}_{k} g_{1k}, \ldots, -\hat{a}_{ k}g_{\ell k}): \hat{a}_k \in \hat{\A}_k\}$. \qed
\end{itemize}

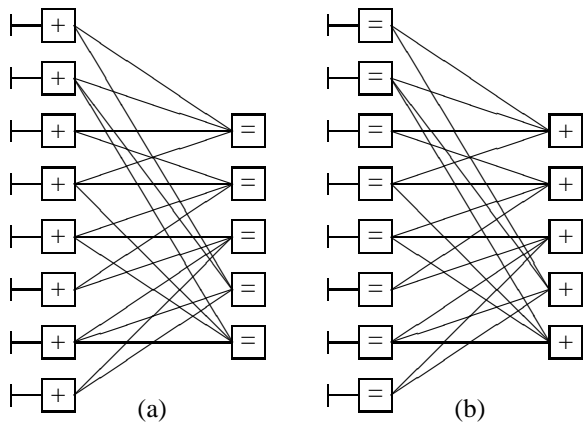
\begin{figure}[h]
\setlength{\unitlength}{4pt}
\centering
\begin{picture}(60,37)(-2,1)
\multiput(2.5,0.5)(0,5){8}{\framebox(3,3){+}}
\multiput(2.5,2)(0,5){8}{\line(-1,0){3}}
\multiput(-0.5,1)(0,5){8}{\line(0,1){2}}
\multiput(20.5,5.5)(0,5){5}{\framebox(3,3){=}}
\put(20.5,27){\line(-3,0){15}}
\put(20.5,27){\line(-3,-1){15}}
\put(20.5,27){\line(-3,1){15}}
\put(20.5,27){\line(-3,2){15}}
\put(20.5,22){\line(-3,1){15}}
\put(20.5,22){\line(-3,-1){15}}
\put(20.5,22){\line(-3,-2){15}}
\put(20.5,22){\line(-3,0){15}}
\put(20.5,17){\line(-3,-1){15}}
\put(20.5,17){\line(-3,0){15}}
\put(20.5,17){\line(-3,-2){15}}
\put(20.5,17){\line(-1,-1){15}}
\put(20.5,12){\line(-3,-2){15}}
\put(20.5,12){\line(-3,-1){15}}
\put(20.5,12){\line(-3,5){15}}
\put(20.5,12){\line(-3,4){15}}
\put(20.5,7){\line(-3,2){15}}
\put(20.5,7){\line(-3,0){15}}
\put(20.5,7){\line(-1,1){15}}
\put(20.5,7){\line(-3,5){15}}
\put(11.5,0){(a)}
\multiput(32.5,0.5)(0,5){8}{\framebox(3,3){=}}
\multiput(32.5,2)(0,5){8}{\line(-1,0){3}}
\multiput(29.5,1)(0,5){8}{\line(0,1){2}}
\multiput(50.5,5.5)(0,5){5}{\framebox(3,3){+}}
\put(50.5,27){\line(-3,0){15}}
\put(50.5,27){\line(-3,-1){15}}
\put(50.5,27){\line(-3,1){15}}
\put(50.5,27){\line(-3,2){15}}
\put(50.5,22){\line(-3,1){15}}
\put(50.5,22){\line(-3,-1){15}}
\put(50.5,22){\line(-3,-2){15}}
\put(50.5,22){\line(-3,0){15}}
\put(50.5,17){\line(-3,-1){15}}
\put(50.5,17){\line(-3,0){15}}
\put(50.5,17){\line(-3,-2){15}}
\put(50.5,17){\line(-1,-1){15}}
\put(50.5,12){\line(-3,-2){15}}
\put(50.5,12){\line(-3,-1){15}}
\put(50.5,12){\line(-3,5){15}}
\put(50.5,12){\line(-3,4){15}}
\put(50.5,7){\line(-3,2){15}}
\put(50.5,7){\line(-3,0){15}}
\put(50.5,7){\line(-1,1){15}}
\put(50.5,7){\line(-3,5){15}}
\put(41.5,0){(b)}
\end{picture}

\caption{(a)  generator realization of $(8,4,4)$ RM code;  (b) parity-check realization of same code.}
\label{RM}
\end{figure}

\vspace{1ex}
\noindent
\textbf{Example 3} (binary Reed-Muller code).  The five binary 8-tuples $11110000, 00111100, 00001111$, $11000011, 01011010$ form a set of five linearly dependent generators for a binary linear block code $\CC$ of length 8 and dimension 4, namely the $(8,4,4)$ first-order Reed-Muller (RM) code. The corresponding generator realization is shown in Fig.\ \ref{RM}(a).

Since $\CC$ is self-dual (\ie $\CC^\perp = \CC$), the same five binary 8-tuples form a set of linearly dependent checks for $\CC$.  Fig.\ \ref{RM}(b) shows the parity-check realization based on these five check 8-tuples, which is the dual to that of Fig.\ \ref{RM}(a).  \qed 

\vspace{1ex}
\noindent
\textbf{Example 2} (cont.).  We note that the primal realization of Fig.\ \ref{Fig2a}(a) is a generator realization based on the three linearly dependent generators $110, 011, 101$, whereas the dual realization  of Fig.\ \ref{Fig2a}(b) is a parity-check realization based on the same three 3-tuples as checks.  \qed \vspace{1ex}

We now settle the observability and controllability properties of such realizations as follows:

\vspace{1ex} 
\noindent
\textbf{Theorem 8 (Observability/controllability of generator and parity-check realizations)}.  A parity-check realization is observable, and a generator realization is controllable.  A generator realization is observable if and only if its generators are linearly independent, and a parity-check realization is controllable if and only if its checks are linearly independent.

\vspace{1ex}
\noindent
\textit{Proof}:  In a parity-check realization, the internal variables are all multiples of external variables;  therefore if all external variables are zero, then all internal variables must be zero.  Thus a parity-check realization is necessarily observable.  By observability/controllability duality, a generator realization must therefore be controllable.

In a generator realization, the internal variables are replicas of the $\ell$ free coefficients, so $\sb \neq \zerob$ if the coefficients are nonzero.  Thus there exists a nonzero configuration $(\zerob, \sb)$--- \ie the realization is unobservable--- if and only if there exists some nontrivial linear combination of the generators that equals the zero codeword $\zerob \in \CC$, which is true if and only if the generators are linearly dependent.  By observability/controllability duality, a parity-check realization is controllable iff its checks are linearly independent.  \qed \vspace{1ex}

It is also interesting to determine controllability using the test of Theorem 6.  In a generator realization, there are $\ell$ equality constraints of dimension 1, and $n$ constraint codes of total dimension $e$, where $e = \dim \SSS$ is the number of one-dimensional internal variables.  Thus  $\sum_i \dim \CC_i = \ell + e$. Since the internal variables are replicas of the $\ell$ free coefficients, the dimension of the behavior $\Bf$ is $\ell$.  Thus $\dim \Bf = \sum_i \dim \CC_i - \dim \SSS$, so a generator realization is controllable.
In a parity-check realization, there are $n$ equality constraints of dimension 1, and $r$ single-parity-check constraints of total dimension $e - r$, where again $e = \dim \SSS$ is the number of internal variables.  We thus have $\sum_i \dim \CC_i = n + e - r$.  By Theorem 8, the realization is observable, so $\dim \Bf =\dim  \CC$.  By Theorem 6, the realization is thus controllable if and only if $\dim \CC = n - r.$

The fact that a parity-check realization is controllable iff its checks are linearly independent nicely illustrates our definition of ``controllable" as ``having independent constraints."

\subsection{Unobservable or uncontrollable $\Rightarrow$ locally reducible}\label{Section 4.5}

We now show how, given an unobservable linear realization of a linear code $\CC$ on a finite graph $\G$, we may reduce the dimension of the behavior $\Bf$ by trimming one state space, without changing the realized code $\CC$.  Thus an unobservable realization is locally reducible.  It follows that the dual uncontrollable realization may be locally reduced by merging the corresponding dual state space.  A similar result was found by Koetter \cite[Lemma 8]{K02}.

For brevity and clarity, we assume that the unobservable realization is linear;  the group case is similar.   We select any state space $\SSS_j$ such that $s_j \neq 0$ in some nonzero configuration $(\zerob, \sb) \in \Bf$.    We choose a basis $\{g_{j\ell} \}$ for $\SSS_j$ with $g_{j1} = s_j$.  The coordinates of $s_j$ in this basis are thus $10\ldots0$.  

Define the subspace $\T_j \subset \SSS_j$ as the set of all states in  $\SSS_j$ whose first coordinate in the chosen basis is 0, and trim the realization by restricting $\Bf$ to the subbehavior $\Bf'$ consisting of those configurations that pass through a state in $\T_j$.  We then replace $\SSS_j$ by $\T_j$, reducing the state space dimension by one.  

The original unobservable configuration $(\zerob, \sb)$ is then not in the trimmed behavior $\Bf'$, since the first coordinate of $s_j$ is $1$.
 However, given any $(\ab, \sb') \in \Bf$, the entire coset $\{(\ab, \sb' + \alpha \sb): \alpha \in \F\}$ is in $\Bf$ and thus realizes $\ab$.  Since $s_{j1} = 1$, there is precisely one element in this coset whose state at time $j$ has first coordinate zero.  Thus the trimmed realization still realizes every $\ab \in \CC$.

In the dual realization, the corresponding local reduction is the merging of the states in $\hat{\SSS}_j$ to their cosets in $\hat{\SSS}_j/(\T_j)^\perp$ via the natural map.  (In the coordinate representation, this amounts to deleting the  first coordinate of $\hat{\SSS}_j$.)  Since the trimmed primal realization still generates $\CC$, the merged dual realization must still generate $\CC^\perp$.
Thus we have proved:

\vspace{1ex}
\noindent
\textbf{Theorem 9}  (\textbf{Local reducibility of unobservable or uncontrollable realizations}). An unobservable linear  realization on a finite graph $\G$ with an unobservable configuration $(\zerob, \sb) \in \Bf$ may be locally reduced by trimming any single state space in the support of $\sb$.  The dual uncontrollable realization may be locally reduced by the dual merging operation. \qed 

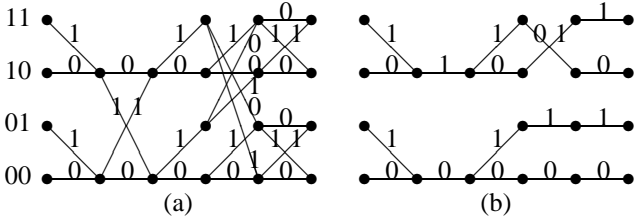
\begin{figure}[h]
\setlength{\unitlength}{4pt}
\centering
\begin{picture}(60,16)(-4, -1)
\put(0,0){\circle*{1}}
\put(-4,-0.5){00}
\put(-4,4.5){01}
\put(-4,9.5){10}
\put(-4,14.5){11}
\put(0,5){\circle*{1}}
\put(0,10){\circle*{1}}
\put(0,15){\circle*{1}}
\put(0,0){\line(1,0){5}}
\put(0,5){\line(1,-1){5}}
\put(2,0){0}
\put(2,3){1}
\put(0,10){\line(1,0){5}}
\put(0,15){\line(1,-1){5}}
\put(2,10){0}
\put(2,13){1}
\put(5,0){\circle*{1}}
\put(5,10){\circle*{1}}
\put(5,10){\line(1,0){5}}
\put(5,0){\line(1,0){5}}
\put(5,10){\line(1,-2){5}}
\put(5,0){\line(1,2){5}}
\put(7,0){0}
\put(7,10){0}
\put(6,6){1}
\put(8,6){1}
\put(10,0){\circle*{1}}
\put(10,10){\circle*{1}}
\put(10,0){\line(1,0){5}}
\put(10,10){\line(1,0){5}}
\put(10,0){\line(1,1){5}}
\put(10,10){\line(1,1){5}}
\put(12,0){0}
\put(12,3){1}
\put(12,10){0}
\put(12,13){1}
\put(15,10){\circle*{1}}
\put(15,15){\circle*{1}}
\put(15,0){\circle*{1}}
\put(15,5){\circle*{1}}
\put(15,0){\line(1,0){5}}
\put(15,5){\line(1,1){5}}
\put(15,0){\line(1,1){5}}
\put(15,5){\line(1,2){5}}
\put(17,0){0}
\put(17,3){1}
\put(19,8){1}
\put(19,12){0}
\put(15,10){\line(1,0){5}}
\put(15,15){\line(1,-3){5}}
\put(15,10){\line(1,1){5}}
\put(15,15){\line(1,-2){5}}
\put(19,10){0}
\put(17,13){1}
\put(19,6){0}
\put(19,1){1}
\put(20,10){\circle*{1}}
\put(20,15){\circle*{1}}
\put(20,0){\circle*{1}}
\put(20,5){\circle*{1}}
\put(20,0){\line(1,0){5}}
\put(20,5){\line(1,0){5}}
\put(20,10){\line(1,0){5}}
\put(20,15){\line(1,0){5}}
\put(22,0){0}
\put(22,5){0}
\put(22,10){0}
\put(22,15){0}
\put(20,0){\line(1,1){5}}
\put(20,5){\line(1,-1){5}}
\put(20,10){\line(1,1){5}}
\put(20,15){\line(1,-1){5}}
\put(21,3){1}
\put(23,3){1}
\put(21,13){1}
\put(23,13){1}
\put(25,10){\circle*{1}}
\put(25,15){\circle*{1}}
\put(25,0){\circle*{1}}
\put(25,5){\circle*{1}}
\put(11,-3){(a)}
\put(30,0){\circle*{1}}
\put(30,5){\circle*{1}}
\put(30,10){\circle*{1}}
\put(30,15){\circle*{1}}
\put(30,0){\line(1,0){5}}
\put(30,5){\line(1,-1){5}}
\put(32,0){0}
\put(32,3){1}
\put(30,10){\line(1,0){5}}
\put(30,15){\line(1,-1){5}}
\put(32,10){0}
\put(32,13){1}
\put(35,0){\circle*{1}}
\put(35,10){\circle*{1}}
\put(35,10){\line(1,0){5}}
\put(35,0){\line(1,0){5}}
\put(37,0){0}
\put(37,10){1}
\put(40,0){\circle*{1}}
\put(40,10){\circle*{1}}
\put(40,0){\line(1,1){5}}
\put(40,10){\line(1,1){5}}
\put(40,0){\line(1,0){5}}
\put(40,10){\line(1,0){5}}
\put(42,0){0}
\put(42,3){1}
\put(42,10){0}
\put(42,13){1}
\put(45,10){\circle*{1}}
\put(45,15){\circle*{1}}
\put(45,0){\circle*{1}}
\put(45,5){\circle*{1}}
\put(45,0){\line(1,0){5}}
\put(45,5){\line(1,0){5}}
\put(45,15){\line(1,-1){5}}
\put(45,10){\line(1,1){5}}
\put(47,0){0}
\put(47,5){1}
\put(46,13){0}
\put(48,13){1}
\put(50,10){\circle*{1}}
\put(50,15){\circle*{1}}
\put(50,0){\circle*{1}}
\put(50,5){\circle*{1}}
\put(50,0){\line(1,0){5}}
\put(50,10){\line(1,0){5}}
\put(52,0){0}
\put(52,10){0}
\put(50,5){\line(1,0){5}}
\put(50,15){\line(1,0){5}}
\put(52,15){1}
\put(52,5){1}
\put(55,10){\circle*{1}}
\put(55,15){\circle*{1}}
\put(55,0){\circle*{1}}
\put(55,5){\circle*{1}}
\put(41,-3){(b)}
\end{picture}
\caption{(a) unobservable tail-biting trellis realization;  (b) uncontrollable dual realization. }
\label{FigU}
\end{figure}

\vspace{1ex}
\noindent
\textbf{Example 4} (tail-biting trellis realization;  from \cite{GLW11, GLW11b} via \cite{GLF}). The linear tail-biting trellis realization shown in Fig.\ \ref{FigU}(a) realizes the binary linear $(5,3)$ block code $\CC = \langle 01110,  00011,  10001 \rangle$.    Since it has a nonzero configuration $(\zerob, \sb)$, this realization is unobservable.  It has five state spaces, which have been coordinatized so that the state values along the nonzero trajectory $(\zerob, \sb)$ are either 10 or 1.  It has five constraint codes $\CC_0 = \langle 01|1|0, 10|0|1 \rangle, \CC_1 = \langle 0|1|1, 1|1|0 \rangle, \CC_2 = \langle 0|1|01, 1|0|10 \rangle, \CC_3 = \langle 00|1|01, 01|1|10, 10|0|10 \rangle$, and $\CC_4 = \langle 00|1|01, 01|1|00,  10|0|10 \rangle$.  This realization is the product of one-dimensional realizations of the four generators $0\underline{111}0, 000\underline{11},$ $\underline{1}000\underline{1}, \underline{01}~\underline{110}$, with the indicated circular spans, and correspondingly its behavior $\Bf$  has the following  four generators:
$$
\begin{array}{cccccccccc}
S_0 & A_0 & S_1 & A_1 & S_2 & A_2 & S_3 & A_3 & S_4 & A_4 \\
\hline
00 & 0 & 0 & 1 & 1 & 1 & 11 & 1 & 00 & 0 \\
00 & 0 & 0 & 0 & 0 & 0 & 00 & 1 & 01 & 1 \\
01 & 1 & 0 & 0 & 0 & 0 & 00 & 0 & 00 & 1 \\
10 & 0 & 1 & 1 & 0 & 1 & 01 & 1 & 10 & 0
\end{array}
$$  

The dual realization to that of Fig.\ \ref{FigU}(a) uses the same variable alphabets and state spaces, and the orthogonal constraint codes $\CC_0^\perp = \langle 01|1|0, 10|0|1 \rangle, \CC_1^\perp = \langle 1|1|1 \rangle, \CC_2^\perp = \langle 0|1|01, 1|0|10 \rangle$, $\CC_3^\perp = \langle 11|0|10, 10|1|11 \rangle$, and $\CC_4^\perp = \langle 10|0|10, 01|1|01 \rangle$, as shown in Fig.\ \ref{FigU}(b).  This dual realization realizes the orthogonal $(5,2)$ code $\CC^\perp = \langle 10111, 01100 \rangle$.  It is a product of realizations of generators $\underline{1}0\underline{111}, \underline{01100}$ with the indicated circular spans, the second span being the entire time axis.  Its behavior $\Bf^\circ$ is correspondingly generated by the following two generators:
$$
\begin{array}{cccccccccc}
S_0 & A_0 & S_1 & A_1 & S_2 & A_2 & S_3 & A_3 & S_4 & A_4 \\
\hline
01 & 1 & 0 & 0 & 0 & 1 & 01 & 1 & 01 & 1 \\
10 & 0 & 1 & 1 & 1 & 1 & 11 & 0 & 10 & 0
\end{array} 
$$
By the test of Theorem 6,  this realization is uncontrollable, since $\dim \Bf^\circ = 2, \sum_i \dim \CC_i^\perp = 9$ and  $\dim \hat{\SSS} = 8$.    (Notice the two disjoint subbehaviors.)

To reduce the unobservable realization of Fig.\ \ref{FigU}(a), let us trim the state space $\SSS_0$.   The resulting constraint codes are then
$\CC_0 = \langle 1|1|0 \rangle, \CC_4 = \langle 01|1|0, 00|1|1 \rangle$; note that both are not trim.

\begin{figure}[h]
\setlength{\unitlength}{4pt}
\centering
\begin{picture}(60,16)(-4, -1)
\put(0,0){\circle*{1}}
\put(0,5){\circle*{1}}
\put(0,0){\line(1,0){5}}
\put(0,5){\line(1,-1){5}}
\put(2,0){0}
\put(2,3){1}
\put(5,0){\circle*{1}}
\put(5,5){\circle*{1}}
\put(5,5){\line(1,0){5}}
\put(5,0){\line(1,0){5}}
\put(5,5){\line(1,-1){5}}
\put(5,0){\line(1,1){5}}
\put(7,0){0}
\put(7,5){0}
\put(6,3){1}
\put(8,3){1}
\put(10,0){\circle*{1}}
\put(10,5){\circle*{1}}
\put(10,0){\line(1,0){5}}
\put(10,5){\line(1,1){5}}
\put(10,0){\line(1,1){5}}
\put(10,5){\line(1,2){5}}
\put(12,0){0}
\put(12,3){1}
\put(12,8){0}
\put(12,11){1}
\put(15,10){\circle*{1}}
\put(15,15){\circle*{1}}
\put(15,0){\circle*{1}}
\put(15,5){\circle*{1}}
\put(15,0){\line(1,0){5}}
\put(15,5){\line(1,1){5}}
\put(15,0){\line(1,1){5}}
\put(15,5){\line(1,2){5}}
\put(17,0){0}
\put(17,3){1}
\put(19,8){1}
\put(19,12){0}
\put(15,10){\line(1,0){5}}
\put(15,15){\line(1,-3){5}}
\put(15,10){\line(1,1){5}}
\put(15,15){\line(1,-2){5}}
\put(19,10){0}
\put(17,13){1}
\put(19,6){0}
\put(19,1){1}
\put(20,10){\circle*{1}}
\put(20,15){\circle*{1}}
\put(20,0){\circle*{1}}
\put(20,5){\circle*{1}}
\put(20,0){\line(1,0){5}}
\put(20,5){\line(1,0){5}}
\put(22,0){0}
\put(22,5){0}
\put(20,0){\line(1,1){5}}
\put(20,5){\line(1,-1){5}}
\put(21,3){1}
\put(23,3){1}
\put(25,0){\circle*{1}}
\put(25,5){\circle*{1}}
\put(11,-3){(a)}
\put(30,0){\circle*{1}}
\put(30,5){\circle*{1}}
\put(30,0){\line(1,0){5}}
\put(30,5){\line(1,-1){5}}
\put(32,0){0}
\put(33,3){0}
\put(30,0){\line(1,1){5}}
\put(30,5){\line(1,0){5}}
\put(32,5){1}
\put(31,3){1}
\put(35,0){\circle*{1}}
\put(35,5){\circle*{1}}
\put(35,5){\line(1,0){5}}
\put(35,0){\line(1,0){5}}
\put(37,0){0}
\put(37,5){1}
\put(40,0){\circle*{1}}
\put(40,5){\circle*{1}}
\put(40,0){\line(1,1){5}}
\put(40,5){\line(1,2){5}}
\put(40,0){\line(1,0){5}}
\put(40,5){\line(1,1){5}}
\put(42,0){0}
\put(42,3){1}
\put(42,8){0}
\put(42,11){1}
\put(45,10){\circle*{1}}
\put(45,15){\circle*{1}}
\put(45,0){\circle*{1}}
\put(45,5){\circle*{1}}
\put(45,0){\line(1,0){5}}
\put(45,5){\line(1,0){5}}
\put(45,15){\line(1,-1){5}}
\put(45,10){\line(1,1){5}}
\put(47,0){0}
\put(47,5){1}
\put(46,13){0}
\put(48,13){1}
\put(50,10){\circle*{1}}
\put(50,15){\circle*{1}}
\put(50,0){\circle*{1}}
\put(50,5){\circle*{1}}
\put(50,0){\line(1,0){5}}
\put(50,10){\line(1,-2){5}}
\put(52,0){0}
\put(53,3){0}
\put(50,5){\line(1,0){5}}
\put(50,15){\line(1,-2){5}}
\put(53,8){1}
\put(52,5){1}
\put(55,0){\circle*{1}}
\put(55,5){\circle*{1}}
\put(41,-3){(b)}
\end{picture}
\caption{State-trimmed realization, and state-merged dual realization.}
\label{FigV}
\end{figure}
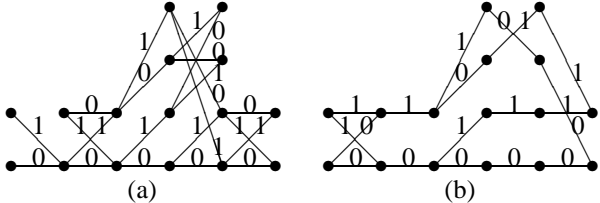

The dual realization merges states in $\hat{\SSS}_0$, yielding the dual constraint codes $\CC_0^\perp = \langle 1|1|0, 0|0|1 \rangle,$ $\CC_4^\perp = \langle 01|1|1, 10|0|0 \rangle$ (both improper).   The resulting merged realization is shown in Fig.\ \ref{FigV}(b);  it is controllable but improper.  

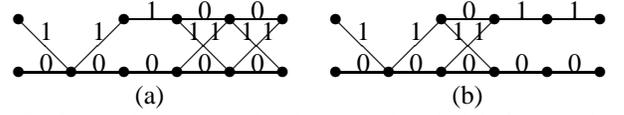
\begin{figure}[h]
\setlength{\unitlength}{4pt}
\centering
\begin{picture}(60,7)(-3, -1)
\put(0,0){\circle*{1}}
\put(0,5){\circle*{1}}
\put(0,0){\line(1,0){5}}
\put(0,5){\line(1,-1){5}}
\put(2,0){0}
\put(2,3){1}
\put(5,0){\circle*{1}}
\put(5,0){\line(1,0){5}}
\put(5,0){\line(1,1){5}}
\put(7,0){0}
\put(7,3){1}
\put(10,0){\circle*{1}}
\put(10,5){\circle*{1}}
\put(10,0){\line(1,0){5}}
\put(10,5){\line(1,0){5}}
\put(12,0){0}
\put(12,5){1}
\put(15,0){\circle*{1}}
\put(15,5){\circle*{1}}
\put(15,0){\line(1,0){5}}
\put(15,5){\line(1,0){5}}
\put(15,0){\line(1,1){5}}
\put(15,5){\line(1,-1){5}}
\put(17,0){0}
\put(16,3){1}
\put(18,3){1}
\put(17,5){0}
\put(20,0){\circle*{1}}
\put(20,5){\circle*{1}}
\put(20,0){\line(1,0){5}}
\put(20,5){\line(1,0){5}}
\put(22,0){0}
\put(22,5){0}
\put(20,0){\line(1,1){5}}
\put(20,5){\line(1,-1){5}}
\put(21,3){1}
\put(23,3){1}
\put(25,0){\circle*{1}}
\put(25,5){\circle*{1}}
\put(11,-3){(a)}
\put(30,0){\circle*{1}}
\put(30,5){\circle*{1}}
\put(30,0){\line(1,0){5}}
\put(30,5){\line(1,-1){5}}
\put(32,0){0}
\put(32,3){1}
\put(35,0){\circle*{1}}
\put(35,0){\line(1,1){5}}
\put(35,0){\line(1,0){5}}
\put(37,0){0}
\put(37,3){1}
\put(40,0){\circle*{1}}
\put(40,5){\circle*{1}}
\put(40,0){\line(1,1){5}}
\put(40,5){\line(1,-1){5}}
\put(40,0){\line(1,0){5}}
\put(40,5){\line(1,0){5}}
\put(42,0){0}
\put(41,3){1}
\put(42,5){0}
\put(43,3){1}
\put(45,0){\circle*{1}}
\put(45,5){\circle*{1}}
\put(45,0){\line(1,0){5}}
\put(45,5){\line(1,0){5}}
\put(47,0){0}
\put(47,5){1}
\put(50,0){\circle*{1}}
\put(50,5){\circle*{1}}
\put(50,0){\line(1,0){5}}
\put(52,0){0}
\put(50,5){\line(1,0){5}}
\put(52,5){1}
\put(55,0){\circle*{1}}
\put(55,5){\circle*{1}}
\put(41,-3){(b)}
\end{picture}
\caption{Trimmed state-trimmed realization and its dual, both minimal.}
\label{FigW}
\end{figure}

If we trim the realization of Fig.\  \ref{FigV}(a), then we obtain a minimal (hence irreducible) realization with state space dimension profile $(1, 0, 1, 1, 1)$, shown in Fig.\ \ref{FigW}(a).  This is a product  realization with generators $0\underline{111}0,  000\underline{11}$, $\underline{1}000\underline{1}$, with the indicated circular spans.

The dual minimal (hence irreducible) realization  is shown in Fig.\ \ref{FigW}(b).   This is a product realization with generators $0\underline{11}00,  \underline{1}0\underline{111}$.  \qed \vspace{1ex}

\subsection{Unobservable and uncontrollable tail-biting trellises}

We now show that the disconnected subtrellises that can be observed in Figs.\ \ref{Fig2a}(b) and \ref{FigU}(b) are characteristic of uncontrollable linear tail-biting trellis realizations.  
This property is reminiscent of the similar classical uncontrollability property for conventional linear state (trellis) realizations.

\vspace{1ex}
\noindent
\textbf{Theorem 10} (\textbf{Uncontrollable tail-biting trellis realizations}).  If a trim linear tail-biting trellis realization is uncontrollable, then its behavior consists of disconnected subbehaviors.

\vspace{1ex}
\noindent
\textit{Proof}.  We first observe that a proper linear tail-biting trellis realization is unobservable if and only if it has a nonzero trajectory $(\zerob, \sb)$ for which $s_i \neq 0$ for all $i$, since properness implies that no constraint code can have a word of the form $(0, 0, s_{i+1})$ or $(s_i, 0, 0)$.
By trim/proper and observability/controllability duality, a trim linear tail-biting trellis realization is therefore uncontrollable if and only if its dual realization has such a nonzero trajectory $(\zerob, \hat{\sb})$.

As in Section \ref{Section 4.5}, we choose a basis $\{\hat{g}_{i\ell} \}$ for each dual state space $\hat{\SSS}_i$ with $\hat{g}_{i1} = \hat{s}_i$.  The coordinates of each $\hat{s}_i$ in this basis are then $10\ldots0$.
Moreover, if we choose a dual basis for each primal state space $\SSS_i$,
then the inner product of elements of $\SSS_i$ and $\hat{\SSS}_i$ is given by the dot product of their coordinate vectors over $\F$.  Thus the set $\hat{s}_i^\perp \subseteq \SSS_i$ of primal states orthogonal to $\hat{s}_i$ is the subspace $\T_i$ of $\SSS_i$ consisting of elements whose first coordinate is zero.  Moreover, the $|\F|$ cosets of $\T_i$ in $\SSS_i$ are the subsets of $\SSS_i$ whose first coordinates are equal to a certain value of $\F$.

Now each dual constraint code $(\CC_i)^\perp$ contains an element $(\hat{s}_i, 0, \hat{s}_{i+1})$ with  $\hat{s}_i$ and $\hat{s}_{i+1}$ having coordinates $10\ldots0$.  It follows that if $(s_i, a_i, s_{i+1})$ is any element of $\CC_i$, then (using the convention that the term $\inner{s_{i+1}}{\hat{s}_{i+1}}$ is negated in the inner product)
\begin{eqnarray*}
0 & = & \inner{(s_i, a_i, s_{i+1})}{(\hat{s}_i, \hat{a}_i, \hat{s}_{i+1})} \\
& = & \inner{s_i}{\hat{s}_i} + \inner{a_i}{\hat{a}_i} - \inner{s_{i+1}}{\hat{s}_{i+1}} \\
& = & s_{i1} - s_{i+1,1}.
\end{eqnarray*}

 It follows that in any trajectory $(\ab, \sb) \in \Bf$, the first coordinates of the state variables must be equal: $s_{i1} = s_{i+1,1}$.  Moreover, since this holds for all $i$, all first state coordinates are equal in all trajectories $(\ab, \sb) \in \Bf$.  Thus the state spaces $\SSS_i$ are partitioned by their first coordinates into $|\F|$ cosets of $\T_i = \hat{s}_i^\perp$, such that state transitions are possible only between cosets with the same first state coordinates.  It follows that the $|\F|$ subbehaviors of $\Bf$ comprising trajectories that have the $|\F|$ different possible values of the first state coordinate are disjoint.  \qed \vspace{1ex}
 
 Using the coset partitioning of this proof, we
define the \emph{zero subbehavior} $\Bf_0$ as the zero coset of $\Bf$, which contains the zero trajectory $(\zerob, \zerob)$. Each of the $|\F|$ cosets of $\Bf_0$ in $\Bf$ then corresponds to the trajectories in $\Bf$ that go through states with a particular value of the first coordinate.

\vspace{1ex}
\noindent
\textbf{Example 2} (cont.).  In the uncontrollable linear tail-biting trellis realization of Fig.\ \ref{Fig2a}(b), the zero subbehavior comprises the all-zero trajectory; the all-one trajectory is its coset. \qed 

\vspace{1ex}
\noindent
\textbf{Example 4} (cont.).  In Fig.\ \ref{FigU}(b), the zero subbehavior comprises the two lower trajectories; the two upper trajectories comprise its coset. \qed \vspace{1ex}

To better understand this result, and to sketch a proof of its converse, we recall the fundamental structure theorem of Koetter and Vardy \cite{KV02, KV03}: every linear state-trim and branch-trim 
tail-biting trellis realization is a product realization.  In other words, the behavior $\Bf$ is the product of $\dim \Bf$ one-dimensional  tail-biting trellis realizations.  Proper realizations are  the product of proper one-dimensional realizations.  Unobservable realizations are those with $\dim \Bf > \dim \CC$ generators.    

One-dimensional realizations may have state support of size $\ell < n$ or size $n$ (called ``degenerate support").  By the controllability test of Theorem 6, the behavior is controllable in the former case (since $\sum_i \dim \CC_i = \ell + 1$, whereas $\sum_i \dim \SSS_i = \ell$ and $\dim \Bf$ = 1), but uncontrollable in the latter (since $\sum_i \dim \CC_i = \sum_i \dim \SSS_i = n$).  It is easy to see that a product of one-dimensional realizations is uncontrollable and has disconnected subbehaviors if and only if a component has degenerate support.

\vspace{1ex}
\noindent
\textbf{Example 4} (cont.).  In Fig.\ \ref{FigU}(b), there is only one possible generator whose state support is not the entire state time axis, namely the generator of the zero subbehavior;  the other generator must be taken from the nonzero coset, and necessarily has degenerate support. \qed \vspace{1ex}

We remark that uncontrollable (but observable) tail-biting trellis realizations in general have too few ``starts" and ``stops"  (transitions between a zero state and a nonzero state);  \eg Fig.\ \ref{FigU}(b) has only one ``start" and ``stop," even though $\dim \CC^\perp = 2$.  Conversely, unobservable (but controllable) tail-biting trellis realizations have too many ``starts" and ``stops;" \eg Fig.\ \ref{FigU}(a) has four ``starts" and ``stops," even though $\dim \CC = 3$.  The difference is the dimension $\dim \SSS^u$ of the dual unobservable subspace in the former case, and $\dim \SSS^u = \dim \Bf - \dim \CC$ in the latter.\footnote{
Koetter and Vardy make the following related observations in \cite[Theorem 4.6 \emph{ff.}]{KV03}.  Define $k = \sum_i \dim \CC_i - \sum_i \dim \SSS_i$.  A linear tail-biting trellis realization that has no degenerate-support generating trajectories (\ie that is controllable, so $k = \dim \Bf$ by Theorem 6) satisfies $k \ge \dim \CC$, with equality if the realization is observable.  If the realization is uncontrollable but observable, then $k < \dim \CC$ (\ie $\sum_i \dim \CC_i - \sum_i \dim \SSS_i < \dim \CC = \dim \Bf$).}

\subsection{General unobservable and uncontrollable realizations}\label{Section 4.7}

In this subsection, we consider a general unobservable linear realization on a finite graph $\G$, and its dual uncontrollable realization.  We will see that uncontrollability does not manifest itself in such an obvious way as in the tail-biting case.

We consider a proper unobservable realization with nonzero configuration $(\zerob, \sb) \in \Bf$.  Then, by properness, the projection $\sb^{(i)}$ of the state configuration $\sb$ on the state alphabet $\SSS^{(i)}$ of any local constraint code $\CC_i$ must have either 0 or 2 or more nonzero components.

We will define a \emph{generalized cycle}\footnote{Interestingly, Chernyak and Chertkov introduce the same concept, which they call a ``generalized loop,"  in their loop calculus \cite{CC06, CC06b}.  They express the partition function of a finite graph as a finite series, whose leading term is a Bethe approximation, and whose higher-order terms correspond to the ``generalized loops" of the graph.} as a subgraph $\G' \subseteq \G$, not necessarily connected, in which all vertices have degree 2 or greater;  \ie there are no isolated (degree-0) or leaf (degree-1) vertices.\footnote{Such a graph is sometimes called a ``2-core;"  however, we avoid this term because a 2-core is implicitly a \emph{maximal} connected subgraph with no leaf vertices.}   Thus the support of a nonzero configuration  $(\zerob, \sb) \in \Bf$ must be a generalized cycle. 

\begin{figure}[h]
\setlength{\unitlength}{5pt}
\centering
\begin{picture}(20,14)(0, 0)
\put(0,0){\framebox(2,2){}}
\put(2,1){\line(1,0){4}}
\put(1,2){\line(0,1){4}}
\put(6,0){\framebox(2,2){}}
\put(8,1){\line(1,0){4}}
\put(12,0){\framebox(2,2){}}
\put(14,1){\line(1,0){4}}
\put(18,0){\framebox(2,2){}}
\put(19,2){\line(0,1){4}}
\put(0,6){\framebox(2,2){}}
\put(2,7){\line(1,0){4}}
\put(1,8){\line(0,1){4}}
\put(6,6){\framebox(2,2){}}
\put(8,7){\line(1,0){4}}
\put(12,6){\framebox(2,2){}}
\put(14,7){\line(1,0){4}}
\put(18,6){\framebox(2,2){}}
\put(19,8){\line(0,1){4}}
\put(0,12){\framebox(2,2){}}
\put(2,13){\line(1,0){4}}
\put(6,12){\framebox(2,2){}}
\put(8,13){\line(1,0){4}}
\put(12,12){\framebox(2,2){}}
\put(14,13){\line(1,0){4}}
\put(18,12){\framebox(2,2){}}
\end{picture}
\caption{Example of a generalized cycle.}
\label{GC}
\end{figure}
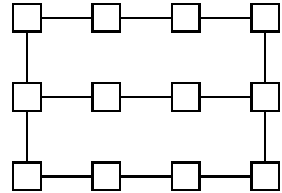

Fig.\ \ref{GC} shows a simple generalized cycle.     In general, a generalized cycle is a union of cycles. 

\begin{figure*}[!t]
\setlength{\unitlength}{5pt}
\centering
\begin{picture}(80,15)(0, -1)
\put(0,0){\framebox(2,2){=}}
\put(2,1){\line(1,0){4}}
\put(1,2){\line(0,1){4}}
\put(6,0){\framebox(2,2){=}}
\put(8,1){\line(1,0){4}}
\put(12,0){\framebox(2,2){=}}
\put(14,1){\line(1,0){4}}
\put(18,0){\framebox(2,2){=}}
\put(19,2){\line(0,1){4}}
\put(0,6){\framebox(2,2){=}}
\put(2,7){\line(1,0){4}}
\put(1,8){\line(0,1){4}}
\put(6,6){\framebox(2,2){=}}
\put(8,7){\line(1,0){4}}
\put(12,6){\framebox(2,2){=}}
\put(14,7){\line(1,0){4}}
\put(18,6){\framebox(2,2){=}}
\put(19,8){\line(0,1){4}}
\put(0,12){\framebox(2,2){=}}
\put(2,13){\line(1,0){4}}
\put(6,12){\framebox(2,2){=}}
\put(8,13){\line(1,0){4}}
\put(12,12){\framebox(2,2){=}}
\put(14,13){\line(1,0){4}}
\put(18,12){\framebox(2,2){=}}
\put(9,-2){(a)}
\put(30,0){\framebox(2,2){+}}
\put(32,1){\line(1,0){4}}
\put(34,1){\circle*{1}}
\put(31,2){\line(0,1){4}}
\put(31,4){\circle*{1}}
\put(36,0){\framebox(2,2){+}}
\put(38,1){\line(1,0){4}}
\put(40,1){\circle*{1}}
\put(42,0){\framebox(2,2){+}}
\put(44,1){\line(1,0){4}}
\put(46,1){\circle*{1}}
\put(48,0){\framebox(2,2){+}}
\put(49,2){\line(0,1){4}}
\put(49,4){\circle*{1}}
\put(30,6){\framebox(2,2){+}}
\put(32,7){\line(1,0){4}}
\put(34,7){\circle*{1}}
\put(31,8){\line(0,1){4}}
\put(31,10){\circle*{1}}
\put(36,6){\framebox(2,2){+}}
\put(38,7){\line(1,0){4}}
\put(40,7){\circle*{1}}
\put(42,6){\framebox(2,2){+}}
\put(44,7){\line(1,0){4}}
\put(46,7){\circle*{1}}
\put(48,6){\framebox(2,2){+}}
\put(49,8){\line(0,1){4}}
\put(49,10){\circle*{1}}
\put(30,12){\framebox(2,2){+}}
\put(32,13){\line(1,0){4}}
\put(34,13){\circle*{1}}
\put(36,12){\framebox(2,2){+}}
\put(38,13){\line(1,0){4}}
\put(40,13){\circle*{1}}
\put(42,12){\framebox(2,2){+}}
\put(44,13){\line(1,0){4}}
\put(46,13){\circle*{1}}
\put(48,12){\framebox(2,2){+}}
\put(39,-2){(b)}
\put(60,0){\framebox(2,2){=}}
\put(62,1){\line(1,0){4}}
\put(61,2){\line(0,1){4}}
\put(66,0){\framebox(2,2){=}}
\put(68,1){\line(1,0){4}}
\put(72,0){\framebox(2,2){=}}
\put(74,1){\line(1,0){4}}
\put(78,0){\framebox(2,2){=}}
\put(79,2){\line(0,1){4}}
\put(60,6){\framebox(2,2){+}}
\put(62,7){\line(1,0){4}}
\put(61,8){\line(0,1){4}}
\put(66,6){\framebox(2,2){=}}
\put(68,7){\line(1,0){4}}
\put(72,6){\framebox(2,2){=}}
\put(74,7){\line(1,0){4}}
\put(78,6){\framebox(2,2){+}}
\put(79,8){\line(0,1){4}}
\put(60,12){\framebox(2,2){=}}
\put(62,13){\line(1,0){4}}
\put(66,12){\framebox(2,2){=}}
\put(68,13){\line(1,0){4}}
\put(72,12){\framebox(2,2){=}}
\put(74,13){\line(1,0){4}}
\put(78,12){\framebox(2,2){=}}
\put(69,-2){(c)}
\end{picture}
\caption{Repetition realization defined on a generalized cycle, its dual, and equivalent dual.}
\label{GCm}
\end{figure*}
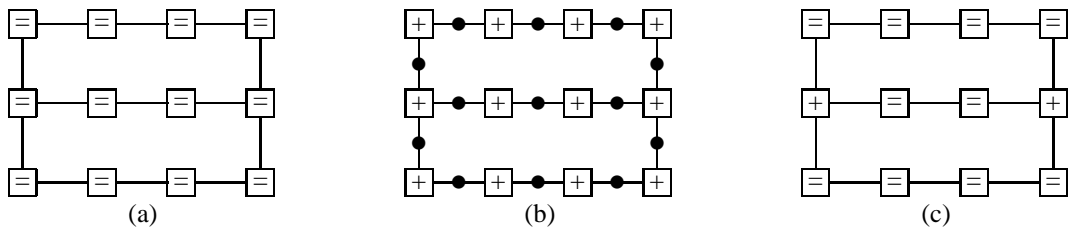

Again, we may choose a basis $\{g_{j\ell} \}$ for each state space $\SSS_j$ in the generalized cycle with $g_{j1} = s_j$, so that every state variable $s_j$ has first coordinate 1 and remaining coordinates equal to 0.  Then the one-dimensional subspace of $\Bf$ generated by the nonzero configuration $(\zerob, \sb)$ is the set of all $(\zerob, \alpha\sb)$ in which the first coordinates of the state vector $\sb$ are all equal to $\alpha \in \F$ on the generalized cycle, and all remaining state coordinates are zero.  Hence the first coordinates are realized by a simple repetition realization $\CC$ over $\F$ defined on the generalized cycle $\G$ in which all constraints are equality constraints, as illustrated in Fig.\ \ref{GCm}(a).

For a realization of the dual uncontrollable code, we choose a dual basis for each dual state space $\hat{\SSS}_j$, so that the inner product of elements of $\SSS_j$ and $\hat{\SSS}_j$ is given by the dot product of their coordinate vectors over $\F$.    Since all configurations $(\hat{\ab}, \hat{\sb}) \in \Bf^\circ$ in the dual realization must be orthogonal to the configuration $(\zerob, \sb)$, it follows that the first coordinates $\sb_1$ in any such configuration must lie in the orthogonal code $\CC^\perp$ to the repetition code $\CC$.  
Thus $\CC^\perp$ is realized by the dual to the repetition realization, in which every equality constraint is replaced by a zero-sum constraint, and a sign inverter (represented by a dot) is inserted in the middle of every edge, as in Fig.\ \ref{GCm}(b).

Such a dual realization may be further simplified as follows.  Since the concatenation of a degree-2 zero-sum constraint and a sign inverter is an equality constraint, and since a linear constraint surrounded by sign inverters is equivalent to the constraint without the sign inverters, we can replace a system such as that of Fig.\ \ref{GCm}(b) by an equivalent system such as that of Fig.\ \ref{GCm}(c).  

In this example, the dual behavior consists of all configurations in which the values of the first state coordinates are constant across the top, middle and bottom chains of degree-2 vertices, say to $\hat{a}_\mathrm{t}, \hat{a}_\mathrm{m}, \hat{a}_\mathrm{b}$, respectively, and furthermore $\hat{a}_\mathrm{t} + \hat{a}_\mathrm{m} + \hat{a}_\mathrm{b} = 0$.  Thus in general 0, 2 or 3 of the values $\hat{a}_\mathrm{t}, \hat{a}_\mathrm{m}, \hat{a}_\mathrm{b}$ are nonzero. The subgraph $\G'' \subseteq \G'$ defined by the nonzero labels is therefore either a set of isolated vertices (\ie the graph comprising all  vertices in $\G'$, but no edges), or a generalized cycle, which can be either one of the three cycles embedded in $\G'$, or the generalized cycle $\G'$ itself.

In the binary case ($\F$ = $\F_2$), a zero-sum codeword must have an even number of ones.  Therefore the support $\G''$ of a nonzero configuration in a dual to a repetition realization must be a generalized cycle in which every vertex has even degree, which is called an \emph{Eulerian graph}.  For example, in the above example,  $\G''$ must be either a set of isolated vertices, or one of the three cycles embedded in $\G'$.  It is easy to see that in general an Eulerian graph is a union of edge-disjoint cycles.

In summary:

\vspace{1ex}
\noindent
\textbf{Theorem 11}  (\textbf{Supports of unobservable and uncontrollable behaviors}). Given a proper unobservable linear  realization on a finite graph $\G$ with an unobservable configuration $(\zerob, \sb) \in \Bf$, the support of $(\zerob, \sb)$ must be a generalized cycle $\G' \subseteq \G$.  A repetition realization as in Fig.\ \ref{GCm}(a) determines the possible values of the first state coordinates for the subspace generated by $(\zerob, \sb)$.  Its dual realization as in Fig.\ \ref{GCm}(b) or (c) determines the possible values of the first dual state coordinates on $\G'$ of all configurations in the behavior $\Bf^\circ$ of the dual uncontrollable realization.  If $\F = \F_2$, then the support of any such dual configuration is an Eulerian graph $\G'' \subseteq \G'$.  \qed \vspace{1ex}

This theorem yields an alternative proof of Theorem 5, since it implies that a proper finite cycle-free linear realization is observable, and thus by duality a trim finite cycle-free linear realization is controllable.  

If a dual constraint code $\CC_i^\perp$ has degree two in the generalized cycle, then it follows from this development that the first coordinates of the two dual state variables involved in $\CC_i^\perp$ must be equal.  Thus, for any chain of degree-2 nodes in a generalized cycle, all first state coordinates must be equal throughout the chain.  Thus when the generalized cycle is simply a single cycle, we retrieve the results of the previous subsection for tail-biting trellis realizations.

\vspace{1ex}
\noindent
\textbf{Example 3} (cont.).  Consider again the unobservable generator realization of the $(8, 4, 4)$ RM code $\CC$ shown in Fig.\ \ref{RM}(a), and its uncontrollable dual parity-check realization shown in Fig.\ \ref{RM}(b).  

\begin{figure}[h]
\setlength{\unitlength}{4pt}
\centering
\begin{picture}(60,37)(-2,1)
\multiput(2.5,0.5)(0,5){8}{\framebox(3,3){=}}
\multiput(20.5,10.5)(0,5){4}{\framebox(3,3){=}}
\put(20.5,27){\line(-3,0){15}}
\put(20.5,27){\line(-3,-1){15}}
\put(20.5,27){\line(-3,1){15}}
\put(20.5,27){\line(-3,2){15}}
\put(20.5,22){\line(-3,1){15}}
\put(20.5,22){\line(-3,-1){15}}
\put(20.5,22){\line(-3,-2){15}}
\put(20.5,22){\line(-3,0){15}}
\put(20.5,17){\line(-3,-1){15}}
\put(20.5,17){\line(-3,0){15}}
\put(20.5,17){\line(-3,-2){15}}
\put(20.5,17){\line(-1,-1){15}}
\put(20.5,12){\line(-3,-2){15}}
\put(20.5,12){\line(-3,-1){15}}
\put(20.5,12){\line(-3,5){15}}
\put(20.5,12){\line(-3,4){15}}
\put(11.5,0){(a)}
\multiput(32.5,0.5)(0,5){8}{\framebox(3,3){+}}
\multiput(50.5,10.5)(0,5){4}{\framebox(3,3){+}}
\put(50.5,27){\line(-3,0){15}}
\put(50.5,27){\line(-3,-1){15}}
\put(50.5,27){\line(-3,1){15}}
\put(50.5,27){\line(-3,2){15}}
\put(50.5,22){\line(-3,1){15}}
\put(50.5,22){\line(-3,-1){15}}
\put(50.5,22){\line(-3,-2){15}}
\put(50.5,22){\line(-3,0){15}}
\put(50.5,17){\line(-3,-1){15}}
\put(50.5,17){\line(-3,0){15}}
\put(50.5,17){\line(-3,-2){15}}
\put(50.5,17){\line(-1,-1){15}}
\put(50.5,12){\line(-3,-2){15}}
\put(50.5,12){\line(-3,-1){15}}
\put(50.5,12){\line(-3,5){15}}
\put(50.5,12){\line(-3,4){15}}
\put(41.5,0){(b)}
\end{picture}

\caption{(a) repetition realization on unobservable support graph $\G'$; (b) dual realization. }
\label{RPCu}
\end{figure}
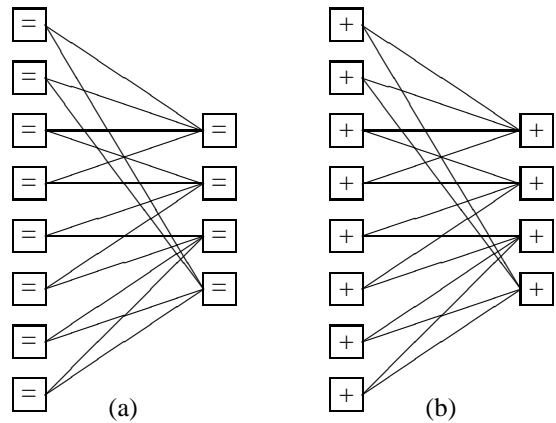

\pagebreak
A repetition realization on the graph $\G'$ supporting the nonzero configuration $(\zerob, \sb)$ of Fig.\ \ref{RM}(a)  is shown in  Fig.\ \ref{RPCu}(a). In this case $\G'$ is itself an Eulerian subgraph, with all vertices having even degree.  

The dual realization on $\G'$ is shown in  Fig.\ \ref{RPCu}(b).  Since all state variables in Fig.\ \ref{RM}(b) are binary, all state configurations must satisfy the constraints of Fig.\ \ref{RPCu}(b).
There are 32 such state configurations, whose supports are the 32 possible Eulerian subgraphs of $\G'$.  The 32 symbol 8-tuples corresponding to these 32 state configurations in Fig.\ \ref{RM}(b) form a linear $(8, 5)$ supercode $\langle 11000000, 00110000, 00001100, 00000011, 01010101 \rangle = \{00, 11\}^4 + \{01, 10\}^4$ of the $(8, 4, 4)$ RM code $\CC$.   The additional constraint imposed by the last equality constraint of Fig.\ \ref{RM}(b) together with these constraints specifies $\CC$.  

We observe that in the uncontrollable realization of Fig.\ \ref{RPCu}(b) there is no partition of the behavior $\Bf$, the state spaces $\SSS_j$, or the constraint codes $\CC_i$ into disjoint, disconnected subsets as in the case of  uncontrollable tail-biting trellis realizations (Theorem 10).  In particular, all state variables are binary and are replicas of symbol variables, so for any pair of state variables that are replicas of different symbol variables, every possible pair of state values occurs in some valid configuration.  Thus the effects of uncontrollability are not as easy to see as in the single-cycle case.   \qed

\subsection{Decoding unobservable or uncontrollable realizations}\label{Section 4.8}

To decode a given linear code $\CC$, we may use any realization of $\CC$ that we like, and any decoding algorithm based on that realization.  In this subsection we will assume that the well-known sum-product (belief propagation) algorithm is used, or any similar iterative message-passing algorithm.  Would there ever be any advantage to using an unobservable or uncontrollable realization of $\CC$?

It is hard to imagine that there would be any advantage to using an unobservable realization, because in any such realization every codeword $\ab \in \CC$ is realized by multiple configurations, and with the sum-product or any similar unbiased algorithm, the weights of all configurations realizing the codeword $\ab$ will be the same.  Thus an iterative decoding algorithm will never converge to a single configuration.

On the other hand, suppose that we use an uncontrollable tail-biting trellis realization of $\CC$.  As we have seen, such a realization partitions into multiple disconnected subrealizations, each realizing a distinct coset of a subcode of $\CC$.  With the sum-product algorithm, each subrealization will be decoded independently of the others.  It is possible that this could actually be helpful, because:
\begin{itemize}
\item  decoding of the ``correct" subrealization (the one containing the transmitted codeword) should be easier, because it realizes a subcode of lower rate than $\CC$ (\ie further from channel capacity); 
 \item decoding of the ``incorrect" subrealizations hopefully should fail in a detectable manner, because all incorrect codewords are far from the correct codeword, so none should be close to the received sequence.
 \end{itemize}
 So if we are able to detect and terminate nonconvergent decoding of the ``incorrect" subrealizations, then we could come out ahead.  
 
 However, similar advantages could be obtained by deliberately choosing any realization of any linear subcode $\CC'$ of $\CC$, using ``translates" of this realization to realize the cosets of $\CC'$ in $\CC$, independently decoding these translates to find the ``best" codeword in each such coset, and then finally comparing these codewords to find the ``best of the best."  The only advantage that we can see of using an uncontrollable tail-biting realization rather than this more general approach is that in the former case a standard sum-product algorithm may be used with little or no modification.
 
 Furthermore, for more general uncontrollable realizations such as the redundant parity-check realization of Example 3, uncontrollability does not in general imply disconnected subrealizations.  Thus we cannot straightforwardly generalize our above argument  for uncontrollable tail-biting realizations.  
 
 In practice, realizations of LDPC codes with redundant parity checks have been considered for various purposes, and shown to be advantageous.  In the original difference-set cyclic codes of Tanner \cite{T81}, redundant checks were included for implementation symmetry;  subsequently, MacKay and Davey \cite{MD01} showed that redundant checks contributed to the good decoding performance of such codes.  More recent theoretical and experimental work (\eg \cite{VK05, KS07, ZSF11}) has shown that redundant checks can reduce the size of the fundamental polytope and therefore the number of pseudocodewords, thus improving performance. 
 
  In summary, there is theoretical and experimental evidence that redundant parity checks can be helpful.  However, we are unaware of any comprehensive study of this question.
  
  \section{Conclusion}
  
For realizations on general finite graphs we have shown:
\[
\textrm{proper}\,\stackrel{\textrm{dual}}{\longleftrightarrow}\,\textrm{trim} \ \textrm{ and }\
\textrm{observable}\,\stackrel{\textrm{dual}}{\longleftrightarrow}\,\textrm{controllable.}
\]
Furthermore, we have shown the following relationships between the various classes of realizations.
Each of the indicated containments is strict.

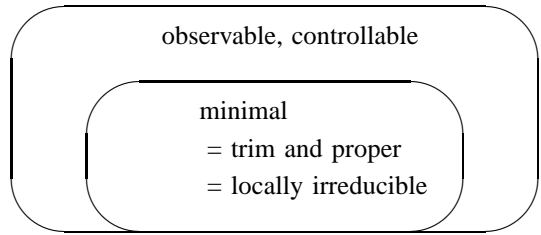
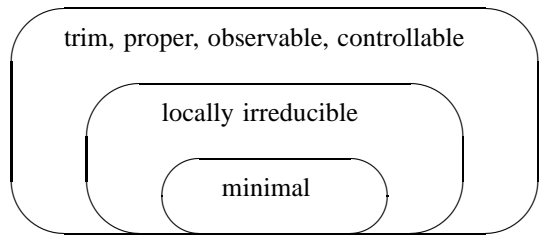
\vspace{2ex}
{
\begin{figure}[h]
\setlength{\unitlength}{1cm}
\centering
\begin{picture}(15,11)(-1,0)
\put(0.8,6){\textbf{General finite graphs}}
\put(3,3){\oval(3,1)}
\put(3,3.5){\oval(5,2)}
\put(3,4){\oval(7,3)}
\put(2.3,3){\textrm{minimal}}
\put(1.5,4){\textrm{locally irreducible}}
\put(0.2,5){\textrm{trim, proper, observable, controllable}}
\put(1,10.5){\textbf{Cycle-free graphs}}
\put(3,8){\oval(5,2)}
\put(3,8.5){\oval(7,3)}
\put(2,8.5){\textrm{minimal}}
\put(2.1,7.5){\textrm{= locally irreducible}}
\put(2.1,8){\textrm{= trim and proper}}
\put(1.5,9.5){\textrm{observable, controllable}}
\end{picture}
\end{figure}
}
\vspace{-15ex}
  
  We have shown that the behavior of uncontrollable tail-biting trellis realizations partitions into disconnected subbehaviors, as with bi-infinite conventional trellis realizations.  However, this phenomenon does not necessarily occur with realizations on more general graphs.  This may be related to the fact that trellis realizations always have product factorizations, whereas general realizations do not.
  
  Finally, we have observed that whereas unobservability seems undesirable in practice, there is evidence that uncontrollability need not hurt, and may even be advantageous.  We recommend further research into this question.
  
  In a subsequent paper \cite{GLF}, we will focus on tail-biting trellis realizations.  We find further criteria for local reducibility, and ultimately give a necessary and sufficient condition for local reducibility of tail-biting trellis realizations, under a more refined definition of local reducibility.
  
  \section*{Acknowledgments}  We are grateful to David Conti, Navin Kashyap, Yongyi Mao and Pascal Vontobel and the reviewers for comments on earlier versions of this paper.

\newpage
\section*{Appendix A:  Minimal cycle-free realizations \\ may have no product factorization}

We now show by a simple counterexample that there is in general no ``product factorization" for minimal cycle-free realizations, in contrast to trellis realizations.  (A more complicated counterexample was given in \cite[Section VII.A]{F03}.)  

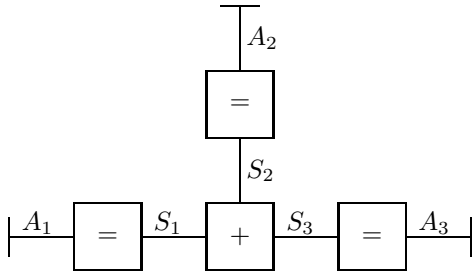
\begin{figure}[h]
\setlength{\unitlength}{5pt}
\centering
\begin{picture}(50,18)(-5, 1)
\put(0,2.5){\line(1,0){5}}
\put(0,1){\line(0,1){3}}
\put(1,3){$A_1$}
\put(5,0){\framebox(5,5){$=$}}
\put(10,2.5){\line(1,0){5}}
\put(11,3){$S_1$}
\put(15,0){\framebox(5,5){$+$}}
\put(17.5,5){\line(0,1){5}}
\put(18,7){$S_2$}
\put(15,10){\framebox(5,5){$=$}}
\put(17.5,15){\line(0,1){5}}
\put(18,17){$A_2$}
\put(16,20){\line(1,0){3}}
\put(20,2.5){\line(1,0){5}}
\put(21,3){$S_3$}
\put(25,0){\framebox(5,5){$=$}}
\put(30,2.5){\line(1,0){5}}
\put(31,3){$A_3$}
\put(35,1){\line(0,1){3}}
\end{picture}
\caption{Cycle-free realization of $(3,2,2)$ binary linear block code.}
\label{Fig1}
\end{figure}

\vspace{1ex}
\noindent
\textbf{Example A} (binary SPC code).
Fig.\ \ref{Fig1} shows a simple minimal cycle-free realization of the $(3,2,2)$ binary linear single-parity-check (SPC) code.  There are three binary-valued symbols, and three binary-valued state spaces.  Three constraints are equality constraints of degree 2, and one constraint is a zero-sum constraint of degree 3.

The full behavior of this realization is generated by the two configurations below:
$$
\begin{array}{cccccc}
A_1 & S_1 & A_2 & S_2 & A_3 & S_3 \\
\hline
1 & 1 & 1 & 1 & 0 & 0 \\
0 & 0 & 1 & 1 & 1 & 1 \\
\end{array}
$$
Thus both generators are ``active" for the state $S_2$, even though its state space dimension is 1.  

By contrast, in a product factorization, the dimension of every state space is equal to the number of generators that are ``active" at that time.  For example, a product factorization would yield a behavior such as the one generated by the two following configurations:
$$
\begin{array}{cccccc}
A_1 & S_1 & A_2 & S_2 & A_3 & S_3 \\
\hline
1 & 1 & 1 & 10 & 0 & 0 \\
0 & 0 & 1 & 01 & 1 & 1 \\
\end{array}
$$
But such a realization is nonminimal, as we see either by comparison with the realization above, or since its top constraint code includes the branch $(a_2, s_2) = (0, 11)$, and is thus improper.

By symmetry, it is clear that a similar situation occurs regardless of which two of the three nonzero configurations are selected as generators.  \qed

\newpage

\section*{Appendix B:  Summary of Koetter \cite{K02}}

We give here a brief summary of some of the main results of Koetter \cite{K02}, using the notation developed in  this paper.  In particular, we represent normal realizations by normal graphs, rather than by ``trellis formations."

As in Section \ref{Section 2.9} of this paper, Koetter (Lemma 1) characterizes the full behavior $\Bf \subseteq \A \times \SSS$ by its orthogonal code $\Bf^\perp \subseteq \hat{\A} \times \hat{\SSS}$.  (In our development, $\Bf^\perp$ is defined as the image of  the homomorphism $\Sigma:  \prod_i {\CC}_i^\perp \to \hat{\A} \times \hat{\SSS}$ that is defined by the sum map $(\hat{\ab}, \hat{\sb}, \hat{\sb}') \mapsto (\hat{\ab}, \hat{\sb} + \hat{\sb}')$;  this is equivalent to Koetter's definition, because for any $j$ at most one of  $\hat{s}_j$ or $\hat{s}'_j$ can be involved in any $(\CC_i)^\perp$.)  The code $\CC$ realized by $\Bf$ is then the projection $\Bf_{|\A}$.

Alternatively, Koetter (Theorem 2) shows that if the orthogonal code $(\Bf^\circ)^\perp \subseteq \A \times \SSS$ to the dual behavior $\Bf^\circ$ is constructed in the same way from $\prod_i \CC_i$, then its cross-section $((\Bf^\circ)^\perp)_{:\A}$ is equal to $\CC$.  In our terms, a simple proof is that by projection/cross-section duality, the orthogonal code to $((\Bf^\circ)^\perp)_{:\A}$ is $(\Bf^\circ)_{|\hat{\A}} = \CC^\perp$.  Koetter gives a direct proof, and then uses this result to prove the normal realization duality theorem (Theorem 3).

In his discussion of vertex merging, Koetter (Lemma 5) first shows that if a constraint code $\CC_i$ is not proper at $\SSS_j$, so $\T = (\CC_i)_{:\SSS_j} \neq \{0\}$, and $\CC_{i'}$ is the other constraint code adjacent to $\SSS_j$, then for any nonzero $t \in \T$ we can add to $\CC_{i'}$ a word whose only nonzero element is $s_j = t$ without changing the code $\CC$ that is realized.  The proof is that $(\Bf^\circ)^\perp$ is unchanged, hence the dual behavior $\Bf^\circ$ is unchanged, hence the dual realization still realizes $\CC^\perp$, hence the primal realization still realizes $\CC$.  This yields a merging procedure equivalent to that shown in our Fig.\ \ref{TPD}(b).

Koetter (Lemma 6) then shows that $(\Bf^\circ)^\perp$ contains a configuration supported by a single state space $\SSS_j$ if and only if the dual realization is not state-trim at $\hat{\SSS}_j$, and that in this case $\SSS_j$ is mergeable.  

In Lemmas 7 and 8 and Theorem 9, Koetter shows that the other condition for mergeability of $\SSS_j$ is that the dual realization is not one-to-one (\ie observable) and contains a nonzero configuration $(\zerob, \hat{\sb}) \in \Bf^\circ$ such that $\hat{s}_j$ is not orthogonal to $\SSS_j$.

Finally, Koetter (Theorem 10) gives a polynomial-time algorithm for identifying mergeable state spaces.

\vspace{4ex}
\section*{Authors' biographical sketches}

	\textbf{G. David Forney, Jr.} received the B.S.E. degree in electrical engineering from Princeton University, Princeton, NJ, in 1961, and the M.S. and Sc.D. degrees in electrical engineering from the Massachusetts Institute of Technology, Cambridge, MA, in 1963 and 1965, respectively.

	From 1965-99 he was with the Codex Corporation, which was acquired by Motorola, Inc. in 1977, and its successor, the Motorola Information Systems Group, Mansfield, MA.  Since 1996, he has been an Adjunct Professor at M.I.T.

	Dr. Forney was Editor of the IEEE Transactions on Information Theory from 1970 to 1973.  He has been a member of the Board of Governors of the IEEE Information Theory Society during 1970-76, 1986-94, and 2004-10, and was President in 1992 and 2008.  He is currently a member of the IEEE Awards Board, and Chair of its Awards Review Committee.  He has been awarded the 1970 IEEE Information Theory Group Prize Paper Award, the 1972 IEEE Browder J. Thompson Memorial Prize Paper Award, the 1990 and 2009 IEEE Donald G. Fink Prize Paper Awards, the 1992 IEEE Edison Medal, the 1995 IEEE Information Theory Society Claude E. Shannon Award, the 1996 Christopher Columbus International Communications Award, and the 1997 Marconi International Fellowship.  In 1998 he received an IT Golden Jubilee Award for Technological Innovation, and two IT Golden Jubilee Paper Awards.  He received an honorary doctorate from EPFL, Lausanne, Switzerland in 2007.  He was elected a Fellow of the IEEE in 1973, a member of the National Academy of Engineering (U.S.A.) in 1983, a Fellow of the American Association for the Advancement of Science in 1993, an honorary member of the Popov Society (Russia) in 1994, a Fellow of the American Academy of Arts and Sciences in 1998, and a member of the National Academy of Sciences (U.S.A.) in 2003.
	
	\vspace{3ex}

	\textbf{Heide Gluesing-Luerssen}
received the Ph.D. degree from the University of Bremen (Germany) in 1991, and the habilitation degree from the University of Oldenburg (Germany) in 2000, both in mathematics.
 
After a postdoctoral fellowship at the University of Bremen from 1991 to 1993, she joined the University of Oldenburg, where she served as faculty member in the Mathematics Department until 2004. From 2004 until 2006 she was a faculty member at the University of Groningen. In 2007 she joined the Department of Mathematics at the University of Kentucky. She has held visiting positions at the University of Notre Dame in 1997--99, at the University of Magdeburg in 2002, and at the University of Kentucky in 2003--04. 

She currently serves as a Corresponding Editor for the {\em SIAM Journal on Control and Optimization\/}, and as an Associate Editor for {\em Advances in Mathematics of Communications}.
Her research interests are in algebraic coding theory and codes on graphs.

\end{document}